\documentclass[fleqn,usenatbib]{mnras}

\usepackage{newtxtext,newtxmath}
\usepackage[normalem]{ulem}

\usepackage[T1]{fontenc}

\DeclareRobustCommand{\VAN}[3]{#2}
\let\VANthebibliography\thebibliography
\def\thebibliography{\DeclareRobustCommand{\VAN}[3]{##3}\VANthebibliography}

\usepackage{graphicx}	
\usepackage{amsmath}	

\newcommand{\Msol}{\mathrm{M_\odot}}
\newcommand{\Mvir}{M_\mathrm{vir}}
\newcommand{\rvir}{R_\mathrm{vir}}

\title[Machine-learning insights into halo profiles]{Insights into the origin of halo mass profiles from machine learning}

\author[Lucie-Smith, Adhikari \& Wechsler]{
Luisa Lucie-Smith,$^{1}$\thanks{E-mail: luisals@mpa-garching.mpg.de}
Susmita Adhikari$^{2, 3, 4}$
and Risa H. Wechsler$^{5,6,7}$
\\
$^{1}$Max-Planck-Institut für Astrophysik, Karl-Schwarzschild-Str. 1, 85748 Garching, Germany\\
$^{2}$Department of Physics, Indian Institute of Science Education and Research, Homi Bhaba Road, Pashan, Pune 411008, India\\
$^{3}$Department of Astronomy and Astrophysics, University of Chicago, Chicago, IL 60637, USA\\
$^{4}$Kavli Institute for Cosmological Physics, University of Chicago, Chicago, IL 60637, USA\\
$^{5}$Department of Physics, Stanford University, 382 Via Pueblo Mall, Stanford, CA 94305, USA\\
$^{6}$Kavli Institute for Particle Astrophysics \& Cosmology, P. O. Box 2450, Stanford University, Stanford, CA 94305, USA\\
$^{7}$SLAC National Accelerator Laboratory, Menlo Park, CA 94025, USA
}

\date{Accepted 2022 June 29. Received 2022 June 28; in original form 2022 May 19}

\pubyear{2022}

\begin{document}
\label{firstpage}
\pagerange{\pageref{firstpage}--\pageref{lastpage}}
\maketitle

\begin{abstract}
The mass distribution of dark matter haloes is the result of the hierarchical growth of initial density perturbations through mass accretion and mergers. We use an interpretable machine-learning framework to provide physical insights into the origin of the spherically-averaged mass profile of dark matter haloes. We train a gradient-boosted-trees algorithm to predict the final mass profiles of cluster-sized haloes, and measure the importance of the different inputs provided to the algorithm. We find two primary scales in the initial conditions (ICs) that impact the final mass profile: the density at approximately the scale of the haloes' Lagrangian patch $R_L$ ($R\sim 0.7\, R_L$) and that in the large-scale environment ($R\sim 1.7~R_L$). The model also identifies three primary time-scales in the halo assembly history that affect the final profile: (i) the formation time of the virialized, collapsed material inside the halo, (ii) the dynamical time, which captures the dynamically unrelaxed, infalling component of the halo over its first orbit, (iii) a third, most recent time-scale, which captures the impact on the outer profile of recent massive merger events. While the inner profile retains memory of the ICs, this information alone is insufficient to yield accurate predictions for the outer profile. As we add information about the haloes' mass accretion history, we find a significant improvement in the predicted profiles at all radii. Our machine-learning framework provides novel insights into the role of the ICs and the mass assembly history in determining the final mass profile of cluster-sized haloes.
\end{abstract}




\section{Introduction}

Dark matter haloes are the fundamental building blocks of cosmic large-scale structure. In our current paradigm for structure formation, haloes formed hierarchically through mass accretion and mergers with smaller structures. The mass accretion history (MAH) of individual haloes can vary significantly; haloes of fixed final mass may have formed through numerous small merger events or a small number of larger, violent mergers. Despite the large variety of non-linear physical processes defining halo formation, $N$-body simulations revealed that the radial density profiles of haloes are fairly well characterized by an $r^{-3}$ decline at large radii and a cuspy profile of the form $r^{-1}$ near the centre. This shape can be described by the isotropic Navarro-Frenk-White (NFW) profile \citep{NFW1997}, a two-parameter functional form given by
\begin{equation}
\rho (r) = \frac{\rho_s}{r/r_s \left( 1 + r/r_s \right)^2},
\label{eq:nfw}
\end{equation}
where $r_s$ is the scale radius, defined as the radius at which $\mathrm{d} \ln \rho/\mathrm{d} \ln r =-2$, and $\rho_s$ is the characteristic density. The scale radius, $r_s$, is often re-written in terms of a concentration parameter, $c \equiv R_\Delta/r_s$, where $R_\Delta$ is the radius containing a mean density of $\Delta$ times the critical (or mean) density of the universe. 

Our understanding of the origin of density profiles is still incomplete. Analytical studies have provided important insights into halo evolution \citep{Gunn:1972sv, FG84, 1985ApJS...58...39B, Ryden:1987ska, Ascasibar:2003mm, 2011ApJ...734..100L}, although several questions still remain unanswered. $N$-body simulations have helped us explore the vast diversity of collapsed structures that emerge from the full distribution of initial perturbations and their complete non-linear evolution. While the assumption of NFW profiles works approximately well, recent work has shown that the density profile near the outskirts can deviate significantly from standard fitting functions, even for stacked ensembles of haloes. In particular, the profiles steepen near what is now known as the splashback radius \citep{Diemer:2014xya, AD14, More:2015ufa, Shi2016, Diemer2022}. The process of virialization that is fundamental to halo relaxation is also not uniformly valid through the entire extent of the halo. Moreover, as haloes grow through mergers and contain significant amount of substructure, the mass tied to subhaloes can also significantly affect the overall profile. For example, \citet{2020MNRAS.499.2426F} shows that spherically-averaged profiles without subhaloes significantly deviate from NFW. Most analytical studies that motivate the origin of the halo profile do not explicitly account for the contribution to the profiles from substructure and their disruption.

Theoretical work on the origin of profiles based on simulations has often focused on investigating statistical correlations between the NFW concentration and aspects of the haloes' assembly history from large population of haloes in $N$-body simulations \citep{Bullock2001, Eke2001, Gao2008, Prada2012, Ludlow2014, Correa2015, Rey2019, WangKuan2020}. These have indicated that the previous mass assembly history of the halo, typically defined as the virial mass of the dark matter halo as a function of time, is deeply connected to its detailed structure of the density profile. In particular, if the haloes' density profiles are parametrized as NFW profiles, late-forming haloes tend to have low concentrations while early-forming haloes have high concentrations \citep{2002ApJ...568...52W}. Beyond studying the correlations between summary parameters like concentration and formation time, there is a large amount of information in the detailed shape of the MAH and the radial mass profiles that can be exploited to get a deeper insight into halo formation. For example, \citet{2013MNRAS.432.1103L} have shown that the shape of the MAH can be used to reconstruct the average inner density profiles ($2 \times r_s$) for haloes in $N$-body simulations, confirming the existence of a deep connection between accretion history and collapsed mass distribution. 

While the MAH specifies the total mass that is enclosed within the virial radius at each time, the distribution of that mass depends on several factors. Once a shell of matter crosses the boundary of a halo, i.e. the splashback radius \citep{Diemer:2014xya, AD14, More:2015ufa, Shi2016}, it encounters a region where matter is in bound orbits; different shells are repeatedly crossing each other at different times \citep{FG84,Ryden88, 1985ApJS...58...39B,  LD10}. Therefore as matter accretes onto the halo, its distribution depends on various factors including the current shape of the matter distribution, the time rate of change of the potential due to mass accretion, the initial angular momentum and the velocity. \cite{2010arXiv1010.2539D} provides a simple toy model to predict the mass distribution from the MAH using symmetry arguments and the conservation of the initial adiabatic invariant of each mass shell set in the initial conditions (ICs). This model was shown to be a good description of an isolated Milky Way mass halo studied using a zoom-in simulation.

Here, we present a machine-learning framework that aims to provide physical insights into the origin of the final mass profile of cluster-mass haloes. We focus on cluster-mass objects, as they are expected to dominate their environment and therefore provide a cleaner sample to study the evolution of structure starting from the ICs to the final collapsed haloes. Our goal is to provide new insights into the role of the ICs and the MAH in determining the final mass distribution of massive dark matter haloes from cosmological $N$-body simulations. Our work overcomes many limitations present in previous analytic works. Firstly, it is not limited to correlations between summary parameters of the MAH and the final profile, such as concentration and formation time; our machine-learning framework works directly on the connection between the full shape of the MAH and that of the final mass profile. Secondly, we do not rely on parameters that are tied to empirical fits to the mass profiles, such as the NFW profile; this allows us to train the network on examples covering the full range of diversity in the mass profiles of individual cluster-mass haloes. Lastly, our work is not limited to averaged profiles that result from stacks of selected `relaxed' haloes; the machine-learning model accurately predicts the mass profile of individual haloes from their MAHs and the ICs. 

We set up two machine-learning frameworks that are trained on two different sets of inputs. The first is trained to predict the final mass profile based solely on the initial density field around the centre-of-mass of each halo's Lagrangian patch. The initial density field is smoothed on a broad range of scales; this gives the algorithm access to information about the local overdensity within the Lagrangian patch, as well as information about the haloes' initial large-scale environment. The second model is given an additional set of inputs describing each halo's MAH. The goal is to interpret the findings of the machine-learning models, thus revealing what the most informative inputs are for predicting the final halo mass profiles. In particular, we measure the importance of the different inputs in predicting the fraction of mass enclosed within different radii, from the inner region out to larger radii approaching the virial radius of the haloes. This allows us to investigate the role of the ICs and the subsequent MAH in determining the final mass distribution of haloes on different scales. 

Our machine-learning framework shares similarity to that used in \citet{Lucie-Smith2018, Lucie-Smith2019}, which revealed new insights into the role of tidal shear effects in the ICs in determining the final mass of dark matter haloes. Their approach consists of training a machine-learning model to infer the final halo mass given different properties of the ICs, and then measuring the importance of the different inputs in predicting the correct final halo mass. In this work, we adopt a similar framework to assess the importance of the ICs and MAH features in predicting the final mass profile of cluster haloes. Recently, \citet{Lucie-Smith2022} extended their work to develop an interpretable deep learning framework to discover the independent set of components required to model the density profiles of dark matter haloes. Here, we focus on a different aspect of dark matter halo profiles: our goal is to improve our understanding of the origin of halo profiles, focusing on how the ICs and haloes' MAH contribute to the build up of the final profiles.

In Sec.~\ref{sec:overview}, we present an overview of our framework, including details on the machine-learning model, the construction of the inputs and outputs, and the training procedure. We present the results for the case where the machine-learning models are trained only on ICs information in sec.~\ref{sec:pred_ics}, and extend the inputs to include the MAH in Sec.~\ref{sec:pred_mah}. We then compare the predictivity of our machine-learning models to linear models in Sec.~\ref{sec:linear}. Finally, we discuss our results in Sec.~\ref{sec:discussion} and present our final conclusions in Sec.~\ref{sec:conclusions}.

\section{Method}
\label{sec:overview}
We trained a machine-learning algorithm to predict the final mass profile of haloes, given different choices of inputs about the initial density field and the MAH. The aim was to provide physical insight into the origin of the mass profile, in particular focusing on understanding how the initial density field and the MAH contribute to the build up of the final halo mass profiles. Using a quantity known as \textit{feature importances}, we investigated which of the inputs that we provide to the algorithm are predominantly responsible in determining the mass enclosed within different radii $r$. Our method differs from previous analytic approaches that focus on statistical, linear correlations between summary statistics of halo profiles and of their accretion histories; the machine-learning algorithm is sensitive to the non-linear causal connection between the inputs and the final halo profile returned by simulations.

\subsection{Simulations}
We used four independent realizations of dark-matter-only $N$-body simulations, three for training and validating the machine-learning model and the other for testing. All simulations were produced using \textsc{L-GADGET} \citep{Springel2001, Springel2005} and a WMAP-5 $\Lambda$CDM cosmological model \citep{Dunkley2009}. The simulations evolve $N=1024^3$ particles in a volume $(1000 \,\mathrm{Mpc}/h)^3$ from $z=99$ to $z=0$. The ICs were generated with different initial random seeds using the publicly available N-GenIC code \citep{Springel2015}.

The haloes were identified at $z=0$ using the \textsc{Rockstar} halo finder \citep{Behroozi2012a}, a phase-space halo-finder that uses an adaptive hierarchical refinement of six-dimensional friends-of-friends and one time dimension to track merged structure over time. Its tree-builder, \textsc{Consistent Trees} \citep{Behroozi2012b} , was also run simultaneously to create halo merger trees. We used \textsc{Rockstar} to identify the halo centres and their virial radii $\rvir{}$. Throughout the paper, we use the virial definition from \citet{BryanNorman1998}, where the virial radius $\rvir{}$ contains a density that is 360 times the critical density at $z=0$.

We restricted our analysis to a small range of massive haloes, $\Mvir{} \in [ 1, 2 ] \times 10^{14} \, \Msol/h$. This allows us to study the build up of the mass profiles of cluster-sized haloes, while minimising the dependency on the total halo mass. On the other hand, the mass range is broad enough to include a large enough population of haloes to construct a representative training set for the machine-learning model. Each simulation contains $\sim 15,000$ haloes in the considered mass range. We test the robustness of our results for a different mass range of cluster haloes, $\Mvir{} \in [ 5, 6 ] \times 10^{13} \, \Msol/h$. However, we do not consider haloes outside the cluster-mass regime; we leave similar studies of lower-mass haloes for future work.

\subsection{Gradient boosted trees}
We used gradient-boosted-trees (GBTs; \citealt{Freund1997, Friedman2001, Friedman2002}), a machine-learning algorithm that combines multiple regression decision trees into a single ensemble estimator. A regression decision tree is a model for predicting the value of a continuous target variable by following a simple set of binary decision rules inferred from the input features. Since individual trees are generally prone to overfitting the training data, they are often combined together to form a more robust ensemble estimator. In GBTs, decision trees are combined via \textit{boosting}: individual trees are added to the ensemble iteratively, such that subsequent trees focus on correcting the prediction errors made by prior trees. Compared to other ensemble estimators, boosting models are typically the most effective at reducing both the bias and the variance contributions to the error in the predictions.

A GBT consists of a sequence of $M$ trees, returning a predicted value $\hat{y} = F_M(\mathbf{x})$, given the inputs $\mathbf{x}$ of a given sample. The accuracy of the model is quantified by the loss function, a measure of how well the model's predictions $\hat{\mathbf{y}}$ fit the ground truth $\mathbf{y}$. We chose a Huber loss function \citep{Huber1965}, given by
\begin{equation}
	\mathcal{L}(y, F_M)= \sum_{i}^{N} \begin{cases}  
	\frac{1}{2} \left(y_{i}-F_M(\mathbf{x}_i)\right)^{2} &\text{for } \left|y_{i}-F_M(\mathbf{x}_i) \right| \leq \alpha, \\ 
	\alpha[\left|y_{i}-F_M(\mathbf{x}_i) \right| - \frac{1}{2}\alpha] &\text{otherwise.}
	\end{cases}
\end{equation}
where $\alpha = 0.9$ and $N$ is the total number of training samples. This function is quadratic for small values and linear for large values of the residual error, thus making it less sensitive to outliers in the data than the commonly-used mean squared error loss. This choice improved the overall accuracy of the model predictions.

During training, the parameters of the model are chosen to minimize the loss function using a gradient-descent optimization scheme. In the context of gradient boosting, the gradient of the loss function with respect to the model's prediction is used as the target variable for each successive tree. More specifically, at any iteration $m$ a new decision tree $f_m(x)$ is added to the existing ensemble $F_{m-1}(\mathbf{x})$ such that the prediction for a given training sample $i$, $F_m(\mathbf{x}_i)$, is updated as
\begin{equation}
    F_m(\mathbf{x}_i) = F_{m-1}(\mathbf{x}_i) + f_m(\mathbf{x}_i),
\end{equation}
where $\mathbf{x}_i$ is the input vector for that training sample. The negative gradient of the loss function with respect to the predictions of the current model $F_{m-1}(\mathbf{x}_i)$ is given by
\begin{equation}
	-\frac{\partial \mathcal{L}(y, F_{m-1})}{\partial F_{m-1}} \Bigg|_{i} =
	\begin{cases}  
	y_{i}-F_{m-1} &\text{for } \left|y_{i}-F_M(\mathbf{x}_i) \right| \leq \alpha, \\ 
	\alpha~\mathrm{sgn}[y_{i}-F_M(\mathbf{x}_i)] &\text{otherwise }.
	\end{cases}
\end{equation}
Thus, the newly added tree $f_m(x)$ is trained to predict the residuals (or, their sign) from the current model based on all prior trees.

The main reason we chose to use GBTs is that, in addition to the predictive power of this algorithm, it also allows for the interpretability of its learning procedure. In fact, an essential aspect of this work lies in determining which features are most informative in predicting the correct final mass profile. We made use of a metric known as \textit{feature importances} \citep{Louppe2013} to measure the relevance of each input feature in training the algorithm to predict the correct target variable. The importance of the $j$-th feature $X_j$ from a single tree $t$ of the ensemble is given by
\begin{equation}
	\mathrm{Imp}_{t}\left(X_j \right)=\sum_{n} \frac{N_{n}}{N_{t}}\left[p -\frac{N_{n_{R}}}{N_{n}} p_{R}-\frac{N_{n_{L}}}{N_{n}} p_{L}\right]
\end{equation}
where $N_t$, $N_n$, $N_{n_{R}}$, $N_{n_{L}}$ are the total number of samples in the tree $t$, at the node $n$, at the right-child node $n_{R}$ and at the left-child node $n_{L}$, respectively. The sum in the equation is over all $n$ nodes where the feature $X_j$ makes a split. The impurity $p$ is given by the choice of splitting criterion, which in our case is the Huber loss function. The final importance of feature $X_j$ given by the ensemble of $T$ trees is the normalized sum over the importances from all trees,
\begin{equation}
	\mathrm{Imp}(X_j)=\frac{\sum_{t=1}^{T} \mathrm{Imp}_{t}\left(X_j\right)}{\sum_{j}^J{\mathrm{Imp}\left(X_j\right)}}.
\label{eq:imp}
\end{equation}

We used \textsc{LightGBM} \citep{Ke2017}, a publicly available gradient boosting framework for machine-learning originally developed by \textsc{Microsoft}. 

\subsection{Machine-learning inputs and outputs}
\begin{figure*}
     \centering
        \includegraphics[width=0.33\textwidth]{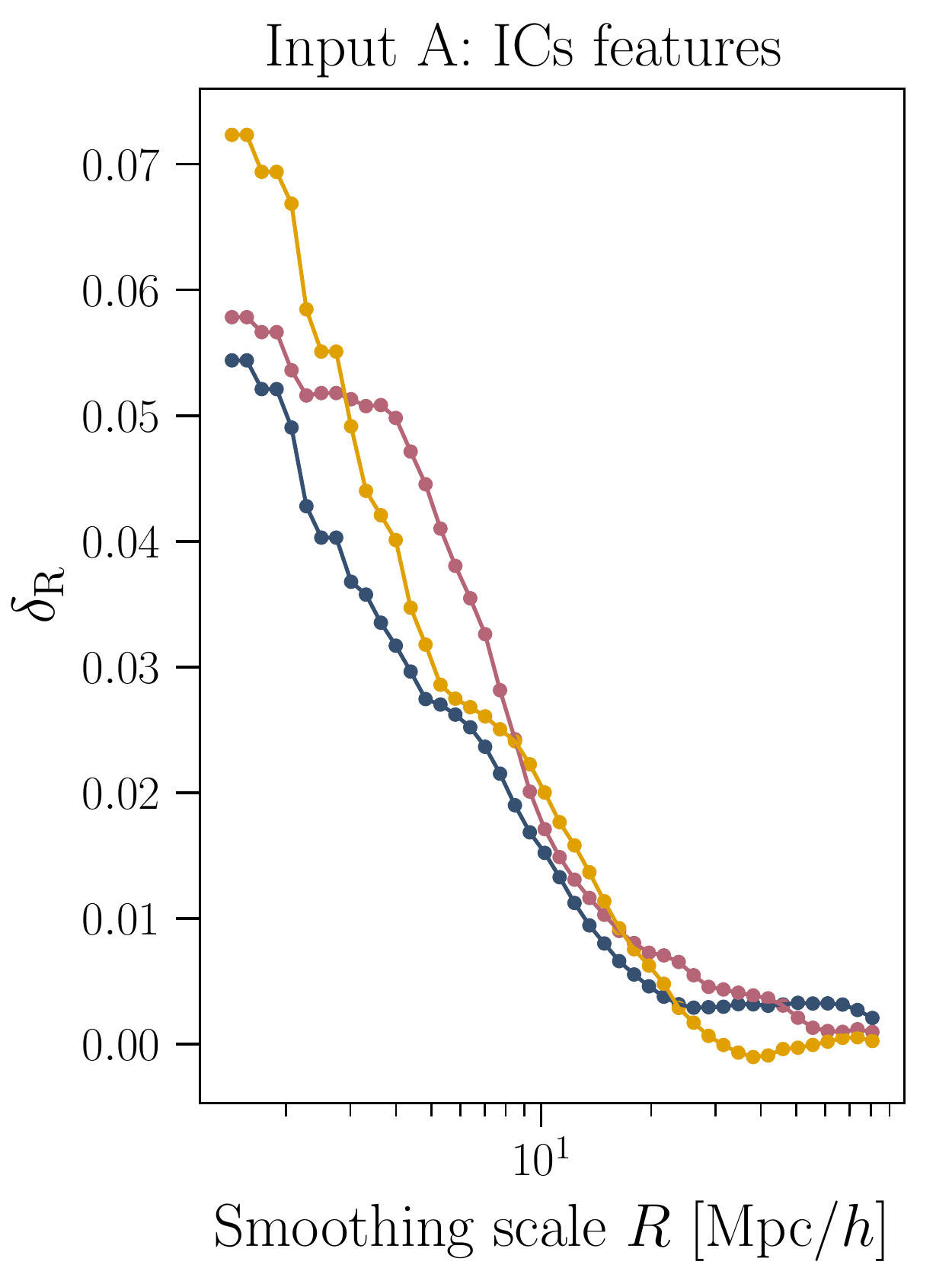}
         \includegraphics[width=0.33\textwidth]{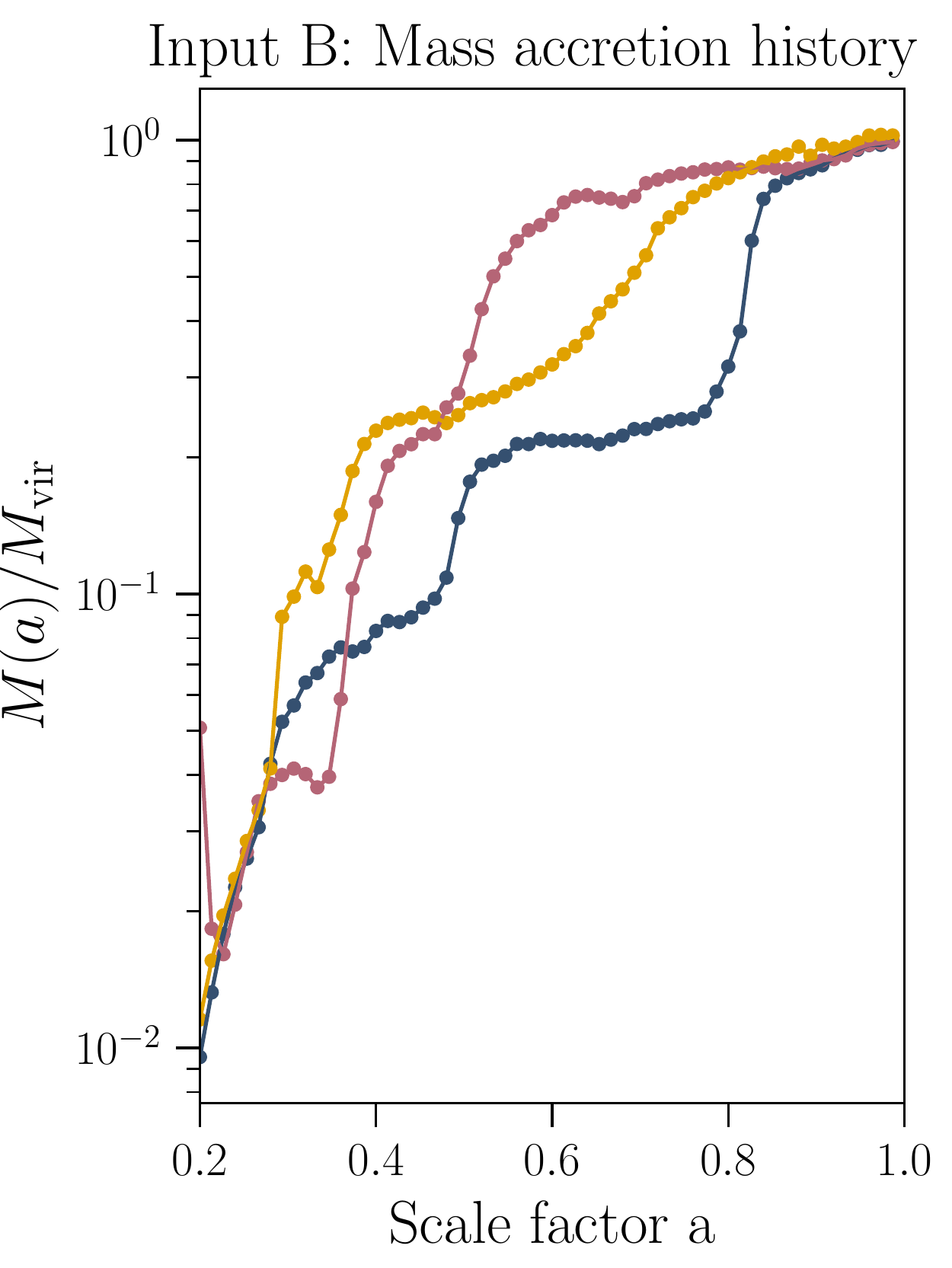}
         \includegraphics[width=0.33\textwidth]{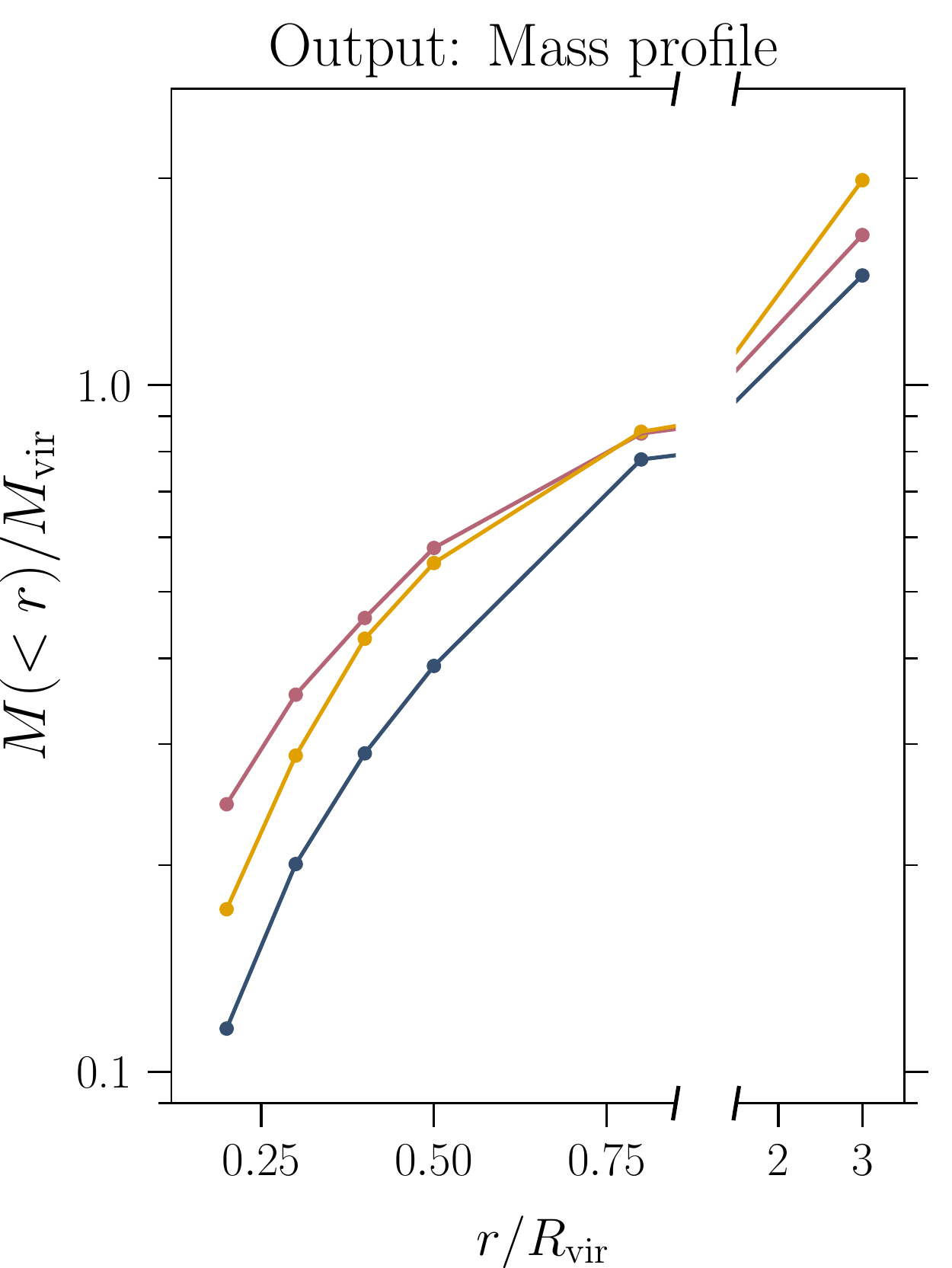}
            \caption{Example of the inputs and output of three dark matter haloes in the training set. The inputs are the ICs features (\textit{left panel}) and the MAH (\textit{middle panel}); the output is the final mass profile (\textit{right panel}). The ICs features are given by the linear density field in the ICs smoothed with a real space top-hat filter on different scales $R$ centred on the centre-of-mass of the haloes' Lagrangian patch. The MAH tracks the mass of the main progenitor of the halo over time. The final mass profile is given by the mass enclosed within different fractions of the virial radius of the halo, normalized by the total virial mass.}
    \label{fig:haloexamples}
\end{figure*}

We considered two sets of inputs, or \textit{features}, for the machine-learning model. The first contains only information about the ICs, whereas the second takes into account information about the assembly history of the halo at later times. The predictive accuracy of the algorithm crucially depends on whether or not the chosen features provide meaningful information to infer the final mass profile of haloes.

\subsubsection{The ICs features}
To construct the ICs features, we computed the initial density contrast field from the positions of particles in the ICs using a cloud-in-cell scheme\footnote{We make use of the Pylian3 libraries, publicly available at https: //github.com/franciscovillaescusa/Pylians3.}. We then smoothed the density contrast $ \delta (\textbf{x}) = \left[ \rho (\textbf{x}) - \bar{\rho}_\mathrm{m} \right]/ \bar{\rho}_\mathrm{m} $, where $ \bar{\rho}_\mathrm{m} $ is the mean matter density of the universe, on a smoothing scale $R$,
\begin{equation}
	\delta (\textbf{x}; R) = \int \delta \left( \textbf{x}^\prime \right) W_{\mathrm{TH}} \left( \textbf{x} - \textbf{x}^\prime; R \right) \text{d}^3 x^\prime,
	\label{smoothed_delta}
\end{equation}
where $W_{\mathrm{TH}} (\textbf{x}, R)$ is a real space top-hat window function 
\begin{equation}
	W_{\mathrm{TH}} (\textbf{x},R) = \begin{cases}  
	\dfrac{3}{4 \pi R^3} &\text{ for } \left| \textbf{x} \right| \leq R, \\ 
	0  &\text{ for }  \left| \textbf{x} \right| >R.
	\end{cases}
\end{equation}

We used 44 logarithmically-spaced smoothing scales in the range $R \in [1.4, 80]~ \mathrm{Mpc}/h$. The window function was centred on the centre-of-mass of each haloes' Lagrangian patch, i.e.  the region of the ICs occupied by the haloes' constituent particles. This choice was made out of three different options that we tried: the centre-of-mass of the Lagrangian patch, the location of the peak of the density smoothed on the size of the Lagrangian patch ($R \sim 7~\mathrm{Mpc}/h$) and the centre-of-mass of the region in the ICs occupied by particles within $0.5\,\rvir{}$ of the final halo. We found that the former, i.e. the centre-of-mass of the Lagrangian patch, yielded a density peak profile with the least flattening at the centre. We therefore present results for this choice throughout the paper. 

To summarise, the ICs features for a given halo consist of the spherically-averaged densities smoothed on 44 different radial scales. Each ICs feature is given by density contrast $\delta_R$ smoothed with a top-hat window function of scale $R$ centred on the centre-of-mass of the Lagrangian patch of the halo. The left panel of Fig.~\ref{fig:haloexamples} shows examples of the ICs features for three different haloes. Since the ICs features consist of spherically-averaged densities, they do not fully capture the impact of large-scale tidal effects in the ICs. These can be included as features by computing the ellipticity and prolateness of the tidal shear tensor smoothed on different scales. We expect tidal effects to have a minimal effect on cluster-sized haloes, although they may become increasingly important for lower-mass haloes. We leave investigating the impact of the tidal field on the final mass profile to future work.

\subsubsection{The MAH features}
The second set of inputs consists of the MAHs of the haloes. We obtain the MAHs from the \textsc{Consistent Trees} algorithm that runs on the output of the \textsc{Rockstar} halo finder. We use the branch of a given halo's history that contains its most massive progenitor and track its mass $M(z)$ over 100 snapshots between $99 < z < 0$, where $z$ is the redshift. We linearly interpolated between the 100 values of $M(a)/\Mvir{}$ and evaluated it at 60 linearly-spaced values of the scale factor in the range $a \in [0.2, 1]$. The middle panel of Fig.~\ref{fig:haloexamples} shows examples of the MAHs for three different haloes.

An alternative, less commonly used way to define the assembly history of haloes is to sum the masses of all collapsed progenitors at any given time \citep{Ludlow2016}. We expect our results to be insensitive to the exact choice of assembly history definition; however, this may become relevant when taking into account different cosmological models, as suggested in \citet{Ludlow2016}.

\subsubsection{The final mass profile output}
The GBT was trained to predict the ratio $M(\leq r)/\Mvir{}$, where $M(\leq r)$ is the mass enclosed within a given radius $r$ from the centre of the halo and $\Mvir{}$ is the virial mass. The mass enclosed within radius $r$ was estimated using all particles within a distance $r$ from the centre of the halo and not just those bound to the halo. We chose $r =  \{ 0.2, 0.3, 0.4, 0.5, 0.8, 3.0 \}\,\rvir{}$. These choices of radii allow us to capture the whole profile, from the inner region to the outer region just before the virial scale. The last value $r = 3\,\rvir{}$ yields the mass enclosed beyond the boundary of the halo, thus probing the environment in which the haloes live at $z=0$. Our framework is not limited to adopting the mass profile as the output, but can be straightforwardly extended to predict halo density profiles instead.

We trained a separate GBT to predict $M(\leq r)/\Mvir{}$ for each radius value. In principle, one could have also trained a single GBT to simultaneously predict the mass enclosed within the different radii. This would not alter the predictive accuracy of the model, which is primarily driven by the information content of the features. On the other hand, training each enclosed mass separately has the advantage of revealing which input features contain relevant information about each $M(\leq r)/\Mvir{}$. Training a single GBT to predict the whole profile simultaneously would have instead combined all that information into a single measure of how important different features are in predicting the mass profile as a whole.

The right panel of Fig.~\ref{fig:haloexamples} shows examples of the final mass profile, normalized by their total virial mass $\Mvir{}$ for three different haloes.

\subsubsection{Training the GBT}
We used all haloes with $\Mvir{} \in [ 1, 2 ] \times 10^{14} \, \Msol/h$ belonging to three simulations to train the GBTs, and those belonging to a different simulation were set aside for validation and testing. The inputs and outputs of the training set haloes were rescaled using a quantile transformation that maps the inputs to a new rescaled variable that follows a Gaussian distribution. This was done to avoid training on features with skewed distributions and outliers. Distributions of this shape were found for the MAH inputs i.e., the values of $M(a)/\Mvir{}$ at different scale factors $a$, as well as the output distributions i.e. $M(\leq r)/\Mvir{}$ at different $r$ values. The quantile transformation provides a robust preprocessing scheme that spreads out the most frequent values, while also reducing the impact of outliers. The transformation was applied on each input feature/output independently. The inputs and outputs of the test set haloes were then also rescaled following the same transformation used for the training set.

The GBT contains hyperparameters which are not optimized via gradient-descent but must be manually fixed prior to training; these include the total number of trees, the learning rate, the tree depth, and the minimum number of samples at any tree leaf. We performed $5$-fold cross validation over a grid-search of values for the hyperparameters to select the best-performing hyperparameters for the GBT. Given a set of values for the hyperparameters, $k$-fold cross validation consists in dividing the training set into k equal-sized sets such that $k-1$ sets are used for training and one is used as a validation set to test the algorithm's performance. The validation set returns a score based on the chosen loss function. The training/validation procedure is repeated k times such that each time a different k set is used for validation and the rest for training. The final score is given by the average score over the k validation sets. This procedure is repeated over a grid-search of values for the hyperparameters, and the set of hyperparameters yielding the best score is retained for the final GBT model. The main benefits of using k-fold cross validation are the following. First, setting aside a subset of the training set for validation ensures that the hyperparameters of the algorithm do not overfit the training data. Second, averaging the score over k validation sets also ensures that the hyperparameters do not overfit any single validation set.

\section{Predicting mass profiles from the ICs}
\label{sec:pred_ics}
\begin{figure*}
    \centering
	\includegraphics[width=\textwidth]{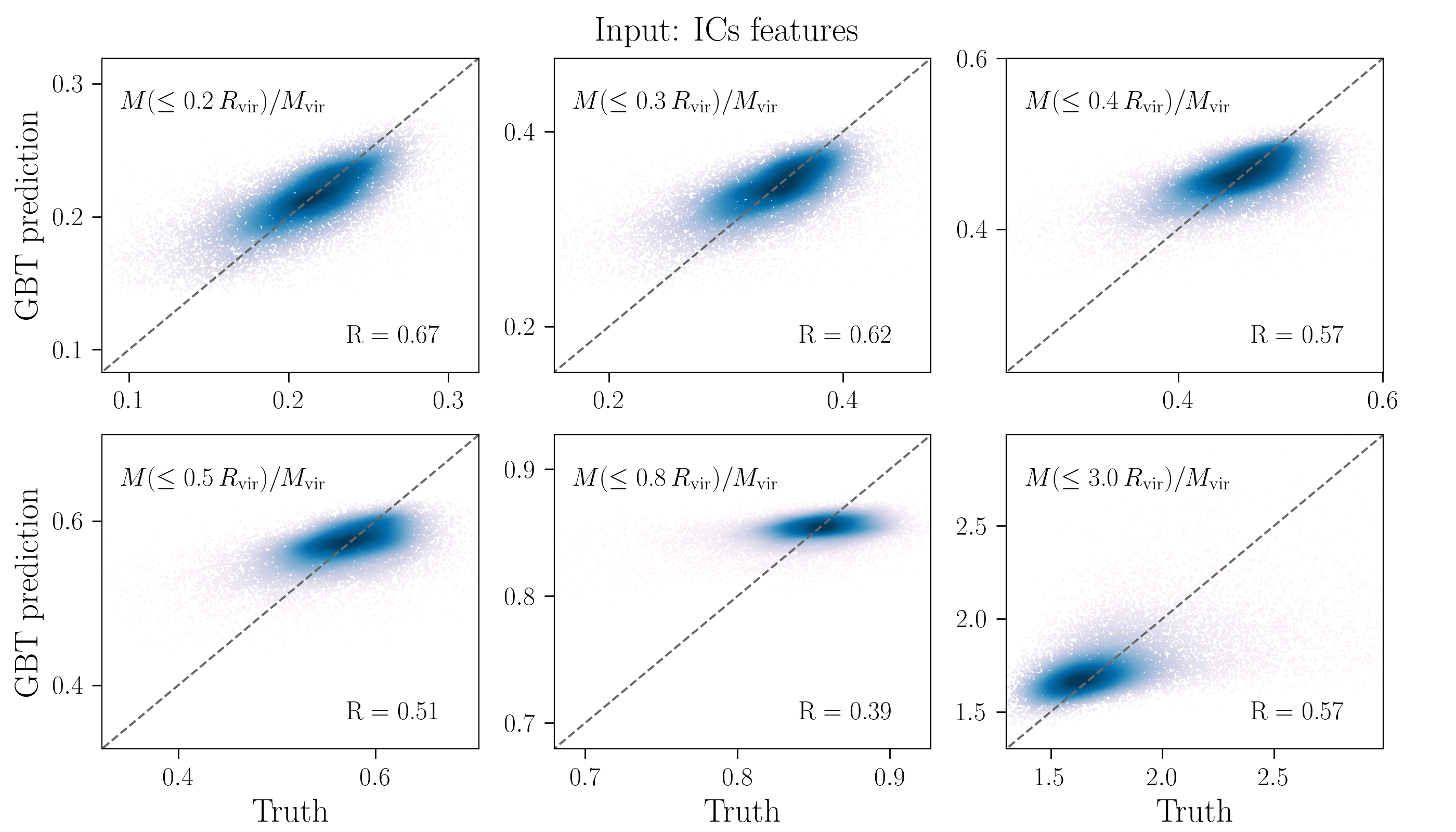}
    \caption{Mass profile predictions when training the GBT only on the ICs features, i.e. smoothed averaged densities around the centre of mass of each halo's Lagrangian patch. Each panel shows the predicted vs true fraction of enclosed mass, $M(\leq r)/\Mvir{}$, for different values of $r = \{ 0.2, 0.3, 0.4, 0.5, 0.8, 3.0\}\,\rvir{}$, for those haloes belonging to the test set. The results are shown as scatter points coloured by their density obtained via 2D kernel density estimation. The value for the correlation coefficient between predicted and ground truth values is shown in the bottom-right corner of each panel. The haloes' inner profile retains memory of the ICs; however, the predictive performance of the GBT decreases as we move towards outer radii, with the exception of $M(\leq 3 \rvir{})/\Mvir{}$.}
    \label{fig:icspredictions}
\end{figure*} 

First, we trained the GBTs using as inputs the ICs features alone. These are the spherically-averaged densities around the centre-of-mass of each halo's Lagrangian region smoothed on different scales. We used six separate GBTs, each trained to predict the mass enclosed within different values of $r$ normalized by the total virial mass. Once trained, the algorithms were used to predict $M(\leq r)/\Mvir{}$ of the haloes belonging to the test set.

Figure \ref{fig:icspredictions} shows the GBT predictions against the simulation's ground truth of $M(\leq r)/\Mvir{}$, for different values of $r$ in each panel. The results are shown as scatter points coloured by their density obtained via 2D kernel density estimation. The top-left panel shows the results for the innermost point in the profile, $M(\leq 0.2\,\rvir{})/\Mvir{}$. We find good agreement between the model's predictions and the simulation ground truth, with a correlation coefficient between the two $R \sim 0.67$. This implies that the inner mass profile retains memory of the density field in the ICs. As we move towards the outer profile, the ICs features become increasingly less predictive. In the case of $M(\leq 0.8\,\rvir{})/\Mvir{}$ (lower-middle panel), the algorithm predicts a very similar value for all haloes in the test set, thus yielding a low correlation coefficient between predicted values and ground truths $R \sim 0.39$. This suggests that there is insufficient information in the initial spherical overdensities for the model to distinguish between haloes that have more or less mass in their outer profile. In Sec.~\ref{sec:pred_mah}, we will show that the addition of informative features to the inputs yields improved mass profile predictions. On the other hand, the ICs features do contain some information about the mass enclosed within $M(\leq 3\,\rvir{})/\Mvir{}$ (bottom-right panel), yielding a correlation coefficient between GBT prediction and ground truth of $\sim 0.57$. Therefore, the environment in which haloes end up at $z=0$ can be inferred from the haloes' density environment in the ICs.

\subsection{Physically interpreting the machine-learning findings}
\begin{figure}
    \centering
	\includegraphics[width=\columnwidth]{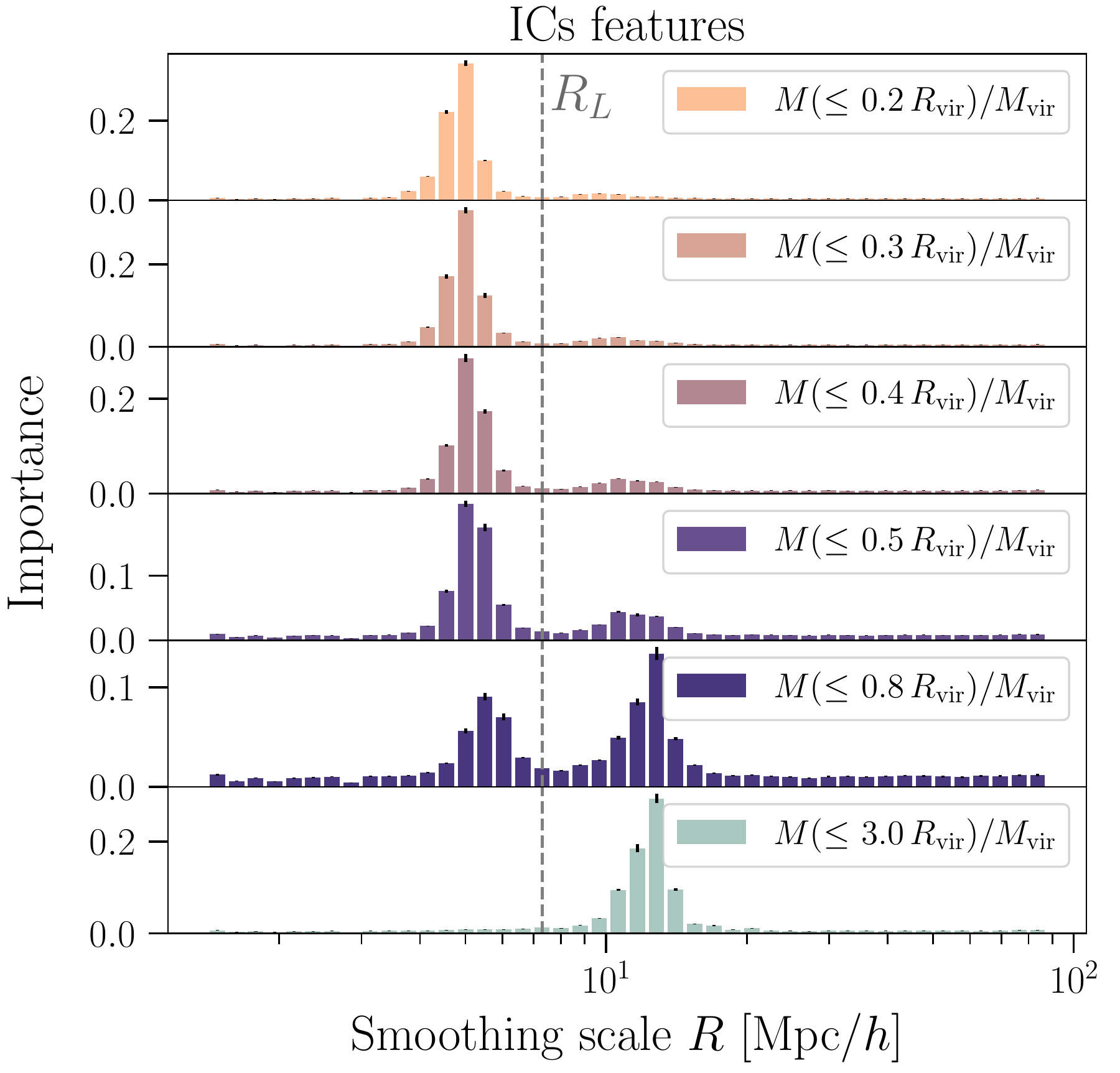}
    \caption{Importances of the ICs density features as a function of their respective smoothing scale $R$. The importance indicates the relevance of each feature in training the model to infer the correct final output. Each panel shows the feature importances when the GBT is trained to infer the mass enclosed within $r = \{ 0.2, 0.3, 0.5, 0.8, 3.0 \}\,\rvir{}$, from top to bottom panel respectively. There are two primary scales in the ICs that impact the final mass profiles of cluster-mass haloes: the density near the Lagrangian radius of the haloes, $R_L$, and that in the large-scale environment.}
    \label{fig:icsimp}
\end{figure}

\begin{figure}
    \centering
	\includegraphics[width=\columnwidth]{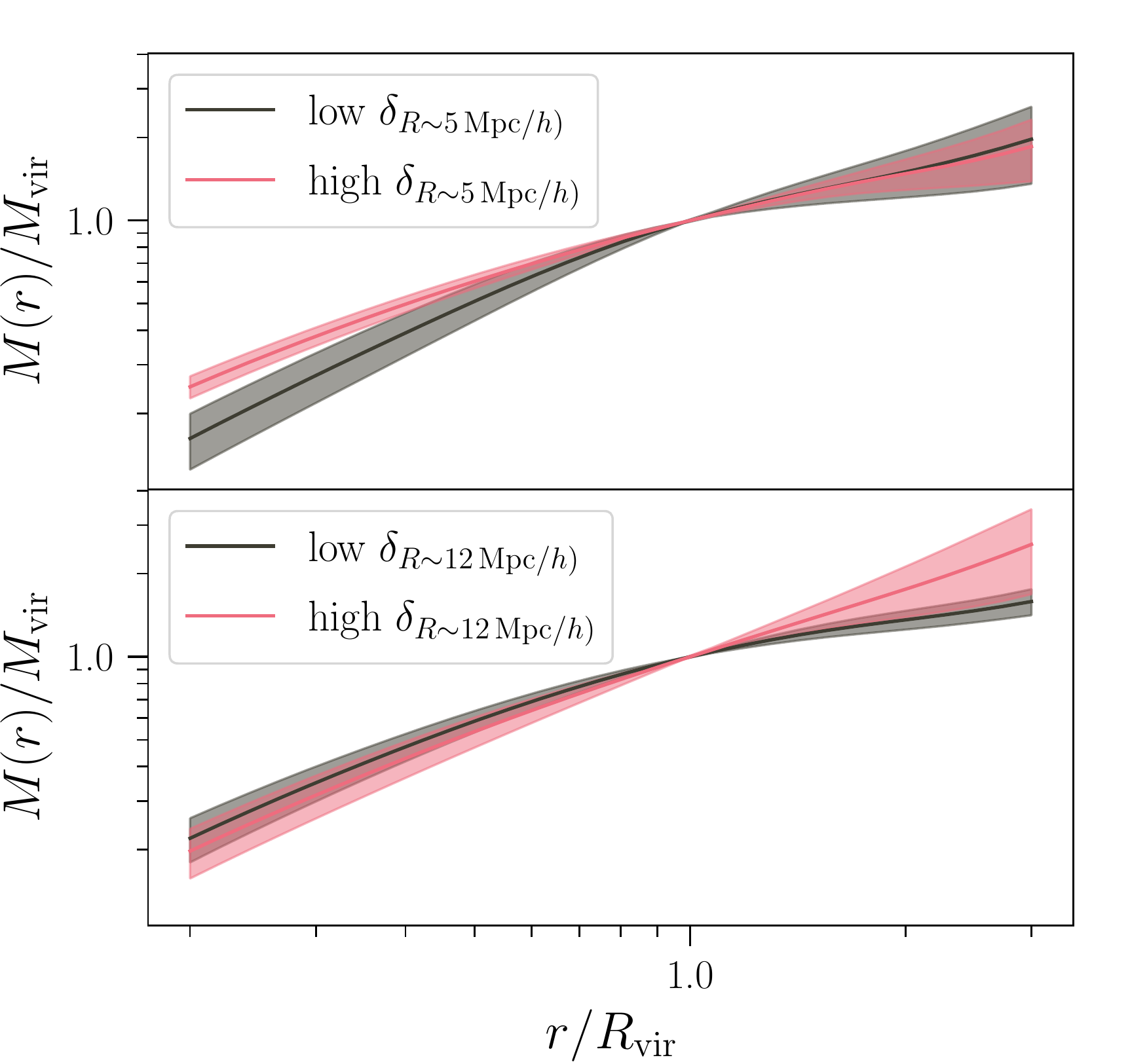}
    \caption{Mean and 1-$\sigma$ errorbands of the mass profiles of two population of haloes: those with low and high values of the density smoothed on $R\sim 5~\mathrm{Mpc}/h$ (upper panel) and that smoothed on $R\sim 12~\mathrm{Mpc}/h$ (lower panel). More specifically, the two populations have values below and above the 80\% confidence interval limits of the distribution of density values of the whole halo population. Higher values of the initial density smoothed on $R\sim 5~\mathrm{Mpc}/h$ lead to more shallow profiles. On the other hand, higher initial large-scale densities ($R\sim 12~\mathrm{Mpc}/h$) yield slightly steeper profiles. Beyond the virial radius, this relation is reversed back to a positive correlation, where higher initial large-scale densities yield higher enclosed masses on large-scales.}
    \label{fig:icssplit}
\end{figure}

\begin{figure*}
    \centering
	\includegraphics[width=\textwidth]{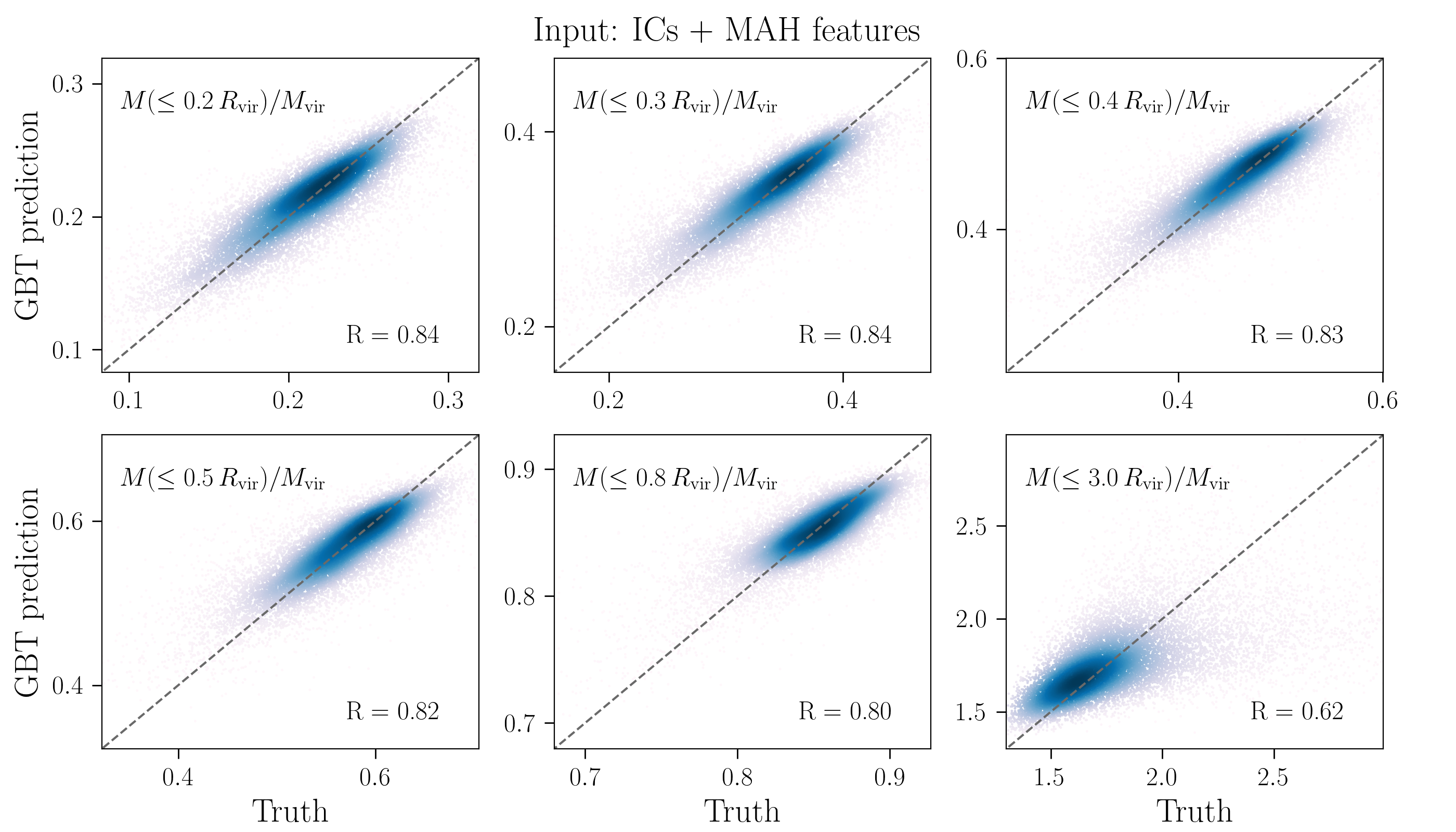}
    \caption{Mass profile predictions when the GBT is trained on both the ICs features and the MAHs of the haloes. Each panel shows the predicted vs true fraction of enclosed mass, $M(\leq r)/\Mvir{}$, for different values of $r = \{ 0.2, 0.3, 0.4, 0.5, 0.8, 3.0\}\,\rvir{}$ (top-left to bottom-right panel). When adding information about the MAHs, the model's predictions of the mass profile up to the virial radius show very good agreement with the ground truths. On the other hand, the MAHs do not affect the predictions for $M(\leq 3\,\rvir{})/\Mvir{}$, which shows similar degree of accuracy as in the bottom-right panel of Fig.~\ref{fig:icspredictions}.}
    \label{fig:icshistpredictions}
\end{figure*}

The algorithm’s performance crucially depends on whether or not the input features contain relevant information about the final mass profile. The fact that the predictions for the inner profile correlate with the ground truths, implies that the former retains memory of its environment in the ICs. By contrast, the ICs features are not able to distinguish between different values of the outer enclosed mass, thus yielding predicted values that only weakly correlate with the ground truth. We now quantify the relevance of each input feature in determining the final outputs using the feature importance measure in Eq.~\eqref{eq:imp}. This allows us to physically interpret the findings of the GBT, revealing where the most important information used by the model to make its predictions is stored in the inputs.

We calculate the relative importances of the ICs features for the GBT models, each trained to predict the mass enclosed within a different fraction of the haloes' virial radius. Fig.~\ref{fig:icsimp} shows the importances of the ICs smoothed densities as a function of their respective smoothing scale R. Each panel (from top to bottom) shows the importances of the features in predicting the mass enclosed within $r = \{ 0.2, 0.3, 0.4, 0.5, 0.8, 3 \}\,\rvir{}$, respectively. The inner profile is most sensitive to the density smoothed on $R\sim 5~\mathrm{Mpc}/h$. This scale can be compared to size of the Lagrangian patch $R_L$, defined as the radius which encloses the total halo mass in the ICs. For haloes of mass $\Mvir{} \in [ 1, 2 ] \times 10^{14} \, \Msol/h$, the median Lagrangian patch size is $R_L \sim 7 \, \mathrm{Mpc}/h$, which closely aligns with the peak scale of the ICs importances ($R\sim0.7~R_L$). This means that the most relevant information to predict the inner profile of haloes lies within the density on scales close to the size of the Lagrangian patch of the haloes. 

As we move towards the outer profile, Fig.~\ref{fig:icsimp} shows that the density at a larger scale of $R\sim 12~\mathrm{Mpc}/h$ becomes progressively more important. In fact, when predicting the mass within $r=0.8\,\rvir{}$, the GBT finds two equally important scales: the density smoothed on $R\sim 5~\mathrm{Mpc}/h$, needed to predict the mass within the inner region of the haloes, and that smoothed on $R\sim 12~\mathrm{Mpc}/h$, needed to predict the outer profile. In other words, we find that the large-scale environment in the ICs affects the outer profile of haloes. This agrees with the work of \citet{Cadiou2021}, who found that the haloes' concentration is strongly affected by the haloes' environment beyond their Lagrangian region in the ICs. 

The bimodality of the feature importances shown in Fig.~\ref{fig:icsimp} is a result of the fact that we are training the model to infer the enclosed mass \textit{normalized} by the total virial mass of each halo, $\Mvir{}$. We checked this by computing the linear correlation between the ICs features and $M(<r)$: we find broad, single-peaked distributions that peak at $r\sim (3 M(<r)/4 \pi \bar{\rho}_\mathrm{m})^{1/3}$, as expected. Instead, the linear correlation between the ICs features and $M(<r)/\Mvir{}$ yields bimodal distributions similar to those of the feature importances in Fig.~\ref{fig:icsimp}.

Figure \ref{fig:icssplit} directly demonstrates the sensitivity of the final mass profiles to the most important ICs features revealed by the machine-learning framework. We considered two populations of haloes: those that have values below and above the $80\%$ confidence interval limits of the distribution of the density smoothed on $R\sim 5~\mathrm{Mpc}/h$ (upper panel) and that smoothed on $R\sim 12~\mathrm{Mpc}/h$ (lower panel). We then computed the mean and 1-sigma error over the (true) mass profiles within each halo population. The upper panel shows that haloes with high values of the initial density smoothed on $R\sim 5~\mathrm{Mpc}/h$ have shallower mass profiles. On the other hand, the lower panel shows that the large-scale initial environment and the profile are anti-correlated: haloes with higher large-scale density at $R\sim 12~\mathrm{Mpc}/h$ in the ICs have slightly steeper mass profiles. This anti-correlation reverts back to a positive correlation at $r>\rvir{}$, meaning that high density values in the ICs imply a high value of enclosed mass far beyond the virial radius $\rvir{}$ at $z=0$.

Our results suggest that there are two primary scales in the ICs that impact the final mass profiles of cluster-mass haloes. The first scale is near the Lagrangian radius of the haloes ($R\sim 5~\mathrm{Mpc}/h \sim 0.7~R_L$) and primarily affects the inner profile; the second is a larger scale ($R\sim 12~\mathrm{Mpc}/h \sim 1.7~R_L$), indicating that the large-scale environment in the ICs contains information about the final outer profile. To test the robustness of our results, we repeated the analysis for a different halo population with mass $\Mvir{} \in [5, 6] \times 10^{13} \, \Msol/h$. The goal was to test whether a different mass range of cluster-sized haloes would also yield two primary scales in the ICs that have a similar relation to $R_L$. We find that the importances of the ICs features peak at $R \sim 4~\mathrm{Mpc}/h \sim 0.7~R_L$ and at $R \sim 8~\mathrm{Mpc}/h \sim~1.6~R_L$, where $R_L$ is the median Lagrangian patch size, $R_L \sim 5.3~\mathrm{Mpc}/h$. This confirms the robustness of our finding that the density near the Lagrangian patch size and in the large-scale environment contain the most relevant information in the ICs to predict the final mass profile of cluster-mass haloes.

\section{The addition of MAH information}
\label{sec:pred_mah}
\begin{figure*}
    \centering
	\includegraphics[width=\textwidth]{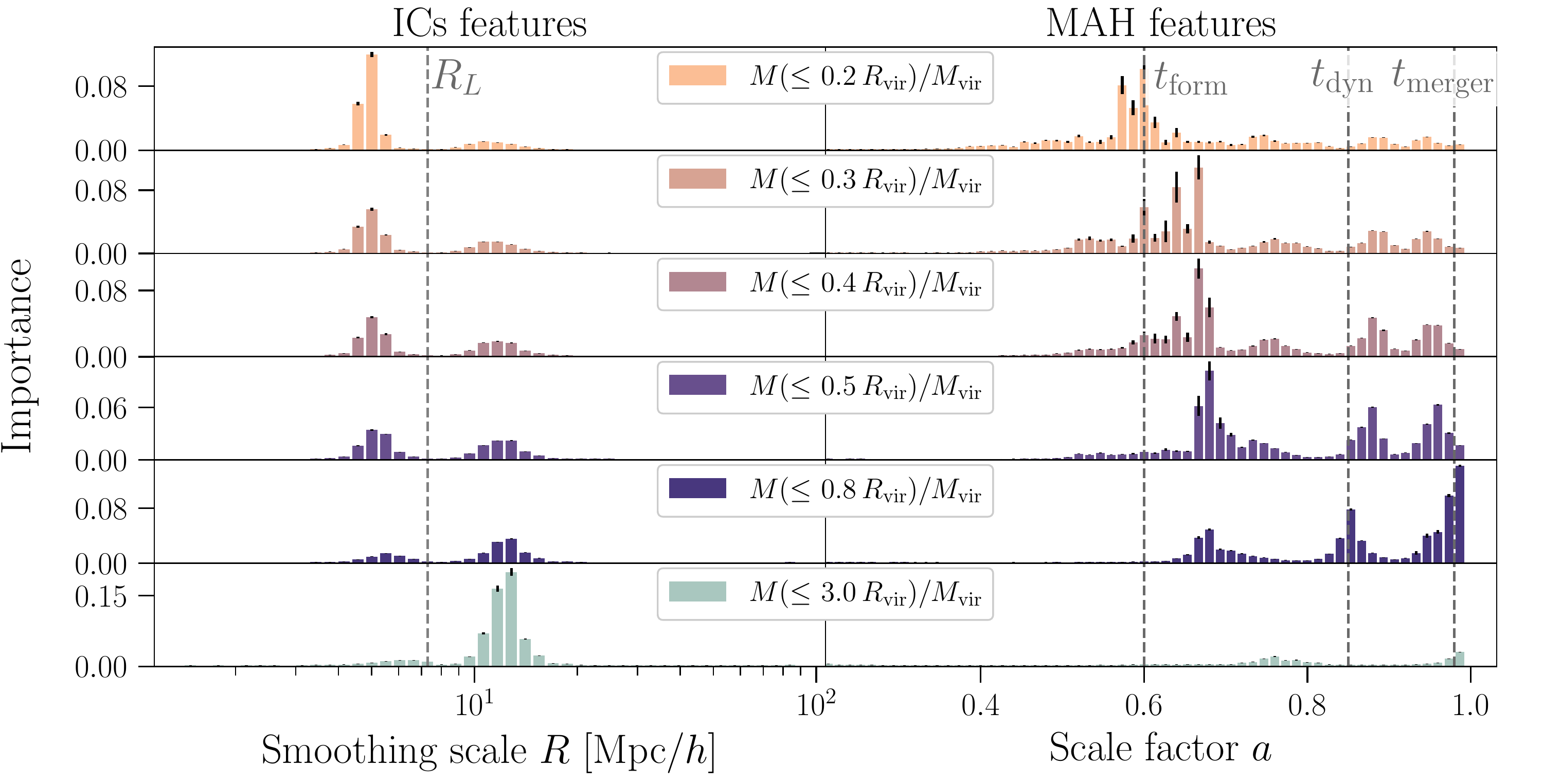}
    \caption{Importances of the ICs and the MAH features used to train GBT models to predict the halo mass profiles. Each panel (from top to bottom) shows the importance of the features when predicting the mass enclosed within $r = \{ 0.2, 0.3, 0.4, 0.5, 0.8, 3.0\}\,\rvir{}$, respectively. Similar to Fig.~\ref{fig:icsimp}, the density near the Lagrangian patch size $R_L$ and the large-scale environment are the two most relevant scales in the ICs to predict the final mass profile of a halo. The MAH importances exhibit three relevant time-scales: the first at $a\sim0.6$ is related to the `formation' time when matter falls into the halo to form the relaxed, virialized component of the halo, the second at $a\sim0.85$ is related to the dynamical time and thus captures dynamically unrelaxed material infalling into the halo, and the most recent times, $a\sim0.95$, related to mergers with massive substructures that have not yet been destroyed. These three time-scales all contribute to the build up of the final profile.
    }
    \label{fig:icshistimp}
\end{figure*}

We now extend the ICs feature set to incorporate additional information on the MAH of the haloes. We study the impact on the predictive performance of the algorithm. This allows us to assess whether the ICs smoothed density and the MAH are able to fully determine the halo mass profiles. Moreover, we assess the interplay between the information stored in the ICs features and that in the MAHs by quantifying the relevance of the features in determining the different parts of the profile.

Figure \ref{fig:icshistpredictions} shows the mass profile predictions from the GBT models, when trained on both the ICs features and the MAHs of the haloes. Similar to Fig.~\ref{fig:icspredictions}, each panel shows the predicted vs true fraction of enclosed mass, $M(\leq r)/\Mvir{}$, for different values of $r$. When adding information about the MAHs, the model's predictions dramatically improve for all values of $r$ within the virial radius. For example, the correlation coefficient for the $M(\leq 0.2\,\rvir{})/\Mvir{}$ case improved from 0.67 to 0.84, and that for the $M(\leq 0.8\,\rvir{})/\Mvir{}$ case improved from 0.39 to 0.8. This shows that the final mass profiles of haloes can be accurately determined by their initial density profile and their subsequent MAH. On the other hand, the MAHs do not strongly affect the predictions for $M(\leq 3\,\rvir{})/\Mvir{}$ (bottom-right panel), whose correlation coefficient with the ground truth changes only from 0.57 to 0.62 with and without MAH information, respectively.

The ICs features of any given halo are themselves correlated with the halo's MAH. This raises the question of whether the ICs are at all needed or if all the relevant information to describe the final mass profiles is already captured by the MAH alone. We tested the performance of the GBT models over the whole mass profile when provided with information about the MAHs alone. We find that the models trained on MAH and ICs features perform better than those trained on MAH alone. When excluding the ICs features, the correlation coefficients between ground truth and predicted values decrease by factors of $3$ -- $6\%$ in all radial bins up to $r=0.8~\rvir{}$, whereas for $M(\leq 3\,\rvir{})/\Mvir{}$ the correlation coefficient decreases by $44\%$. This suggests that the ICs features carry additional information to that already contained in the MAHs.

\subsection{The importances of the ICs and MAH features}
\label{sec:impmahs}
Figure \ref{fig:icshistimp} shows the importances of all features: the ICs densities as a function of their smoothing scale R (left columns) and the MAHs i.e. $M(a)$ as a function of scale factor $a$ (right columns). Each horizontal panel shows the relevance of the features in determining the different enclosed masses, from the inner mass profile to the outer profile ending with the mass enclosed within $3\times$ the virial radius (top to bottom panels). The importances of the ICs features exhibit similar trends to those in Fig.~\ref{fig:icsimp}. The density smoothed on $R\sim 5~\mathrm{Mpc}/h$ carries important information about the inner profile; as we move towards larger radii, the density smoothed on larger scales, i.e. $R\sim 12~\mathrm{Mpc}/h$, becomes increasingly important. We find that the ICs information is as important as the early mass assembly around $a\sim 0.6$ in establishing the haloes' inner profile. On the other hand, the information contained in the ICs about the outer profile becomes subdominant over information about the assembly history at late times. This explains the large improvement in the accuracy of the predictions of the outer profile ($r=0.8\,\rvir{}$) as we provide the algorithm with information about the mass assembly history.  

The mass enclosed within $r=3\,\rvir{}$ shows the opposite trend: the information in the MAH is largely irrelevant and only the large-scale density in the ICs contains important information. That is expected since the specific assembly history of a halo does not significantly impact the large-scale environment in which the halo is embedded.

We find that the MAH importances peak at three different time-scales. The first important time-scale found by the GBT is at early times ($a\sim 0.6$) and it affects the final mass profile at all radii. This peak shifts gradually to later times for larger radii: the inner profile is affected by the early-time assembly, whereas the outer profile is more sensitive to the later time mass assembly. This result is broadly consistent with existing work, showing that the inner profile is established in the early fast-accretion phase, whereas the outer profile is dominated by the slow-accretion phase of mass on to an existing central object at later times \citep{Wechsler2002, Zhao2003a, Zhao2003b, Li2007, Lu2006}. The next two important time-scales in the MAH of the haloes are found at $a\sim 0.85$ and $a\sim 0.98$; these two time-scales become increasingly important as we move towards the outer profile. 

We provide a more detailed physical interpretation of the three most relevant time-scales in the MAHs as follows.

\subsubsection{The first peak in the MAH importances: halo formation time}
\label{sec:peakone}
We interpret the first peak as the halo `formation time': the approximate virialization time-scale of collapsed material at a given radius. This peak corresponds to the phase of early accretion when there is fast accretion onto the halo that forms the relaxed, virialized component of the halo at any given radius $r$ \citep{Zhao2003a}.  This peak gradually shifts towards later times as we move towards the outer profile; this reflects the fact that haloes first form their inner region and gradually accrete mass in the outskirts over time and also the fact that orbits of particles shrink with time. The mean formation time\footnote{We define halo formation time as the time at which a halo has acquired half of its final virial mass.} of the haloes in our sample is $a\sim 0.6$, which approximately corresponds to the time-scale of the first peak in the MAH importances.

\begin{figure}
    \centering
	\includegraphics[width=\columnwidth]{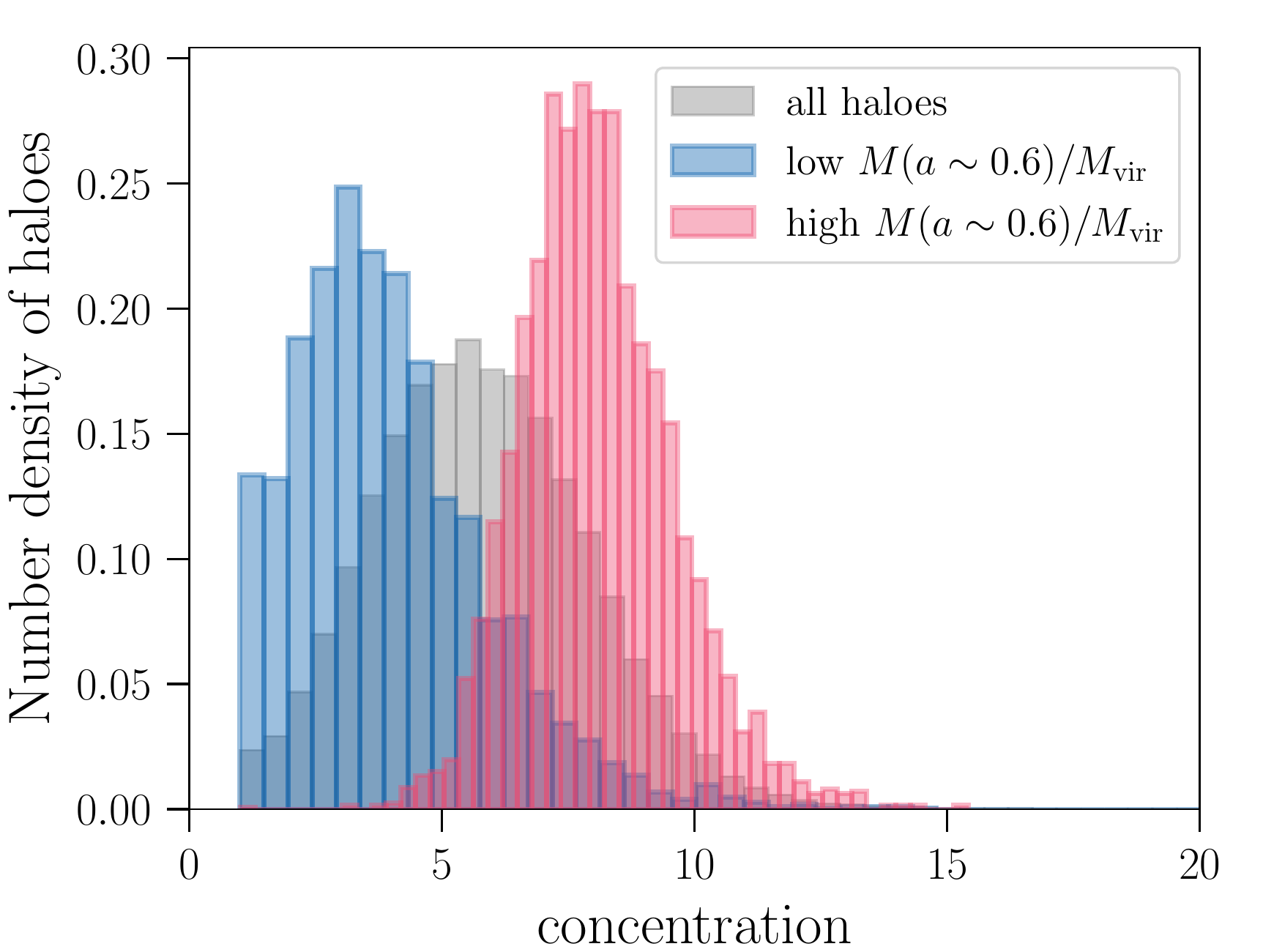}
    \caption{Distribution of concentration values for two populations of haloes that have values above and below the 80\% confidence interval limits of the distribution of $M(a\sim0.6)/\Mvir{}$, respectively. $M(a\sim0.6)/\Mvir{}$ is the earliest most important feature in the haloes' mass assembly history. The grey histograms shows the distribution of concentration values for the entire halo population. High (low) values of $M(a\sim0.6)/\Mvir{}$ lead to more (less) concentrated haloes. This trend reflects the well-known correlation between early-forming and high-concentrated haloes, thus confirming the relation between halo formation time and the first peak in the MAH importances.}
    \label{fig:splithist_imp}
\end{figure}

Existing numerical studies have shown that halo formation time affects the final collapsed density distribution of haloes \citep{Wechsler2002, Ludlow2014}. In particular, it has been shown that late-forming haloes tend to have low values of the NFW concentration while early-forming haloes have high NFW concentration \citep{Wechsler2002}. Our work directly connects the MAH to the fraction of enclosed mass at different radii $M(<r)/\Mvir{}$, without the need to rely on summary parameters such as the concentration. To test our physical interpretation that the first importance peak is related to halo formation time, we also investigate the connection between the first importance peak and concentration. To do so, we split the haloes in two populations based on their value of $M(a\sim 0.6)/\Mvir{}$. The two populations consist of haloes with $M(a\sim 0.6)/\Mvir{}$ below and above the $70$\% confidence interval limits of the distribution of $M(a\sim 0.6)/\Mvir{}$ values over the entire halo population. Fig.~\ref{fig:splithist_imp} shows the concentration of the entire halo population (grey histogram) and that of the two halo populations with low and high values of $M(a\sim 0.6)/\Mvir{}$. We find that the two populations have two distinct concentration values: haloes with high (low) values of $M(a\sim 0.6)/\Mvir{}$ are more (less) concentrated. This result confirms that the first importance peak is a signature of the well-known link between halo profiles and formation time. In Appendix \ref{sec:conc}, we split haloes by their concentration into two populations, and compare the predictions for models that are trained separately on the two halo populations.

Similar to the ICs feature importances case, we compared the MAH importance peaks to those that result from training (and testing) on a lower-mass halo population with $\Mvir{} \in [5, 6] \times 10^{13} \, \Msol/h$. We find that the first peak in the MAH importances shifts slightly to earlier times ($\Delta a \sim 0.03$), reflecting the fact that less massive halos form earlier. This result further confirms the link between the first peak in the MAH importances and formation time.

\subsubsection{The second peak in the MAH importances: dynamical time}
\label{sec:peaktwo}

The second most important time in the MAHs of the haloes lies in the range $a\sim 0.85- 0.9$, or lookback time $t \sim 1.2 - 2.2$ Gyrs; this is likely related to the dynamical time of cluster-sized haloes. The dynamical time approximately describes the time required for material to cross the halo moving at the typical infall velocity of the material. For cluster-sized haloes of $\Mvir{} \sim 10^{14} \, \Msol{}/h$, $\rvir{} \sim 1 \, \mathrm{Mpc}/h$ and $v_{\mathrm{infall}} \approx 1000 \, \mathrm{km/s}$, the time required for material to cross the entire halo is $t_\mathrm{cross} = 2 \rvir{}/v_\mathrm{infall} = 2.8$ Gyrs. On the other hand, the pericenter crossing time is $t_\mathrm{per} = \rvir{}/v_\mathrm{infall} = 1.4$ Gyr. The second peak in the MAHs importances lies within these timescales, suggesting that the GBT captures the impact of the dynamically unrelaxed, infalling material on the halo mass profile.

\subsubsection{The third peak in the MAH importances: mergers}
\label{sec:peakthree}

The third, most recent time-scale at $a\sim 0.98$ becomes increasingly relevant as we move towards the outer profile, similar to the second peak at $a\sim 0.85$. We will demonstrate in Sec.~\ref{sec:subhaloes} that this peak captures the impact of massive substructures that have recently merged with the haloes and that have not yet had time to be destroyed. Recent mergers with substructures mostly affect the outer profile; therefore, information about the most recent assembly history is required in order to correctly determine the mass enclosed within the outskirts of haloes.

When compared to the MAH importances for the lower-mass halo population with $\Mvir{} \in [5, 6] \times 10^{13} \, \Msol/h$, we find that both the second and third most relevant MAH features peak at similar timescales. This is expected, as the difference in the dynamical time of the two halo mass populations is negligible.

\subsection{The impact of subhaloes on the MAH importances}
\label{sec:subhaloes}
Clusters haloes, being massive, are usually in the densest environments. They are actively accreting other massive subhaloes, and because they are relatively young a significant fraction of the mass is made of massive substructure (even $M_{\rm sub} >0.01 M_{\rm host}$) that has not been tidally destroyed due to the multiple orbital passage or dynamical friction.

We investigate the change in the feature importances if we consider only the diffuse component of dark matter within clusters when constructing the mass profiles. To do this, we removed all particles that belong to subhaloes. Following \citet{2020MNRAS.499.2426F}, we removed subhaloes particles using the Particle Tables provided by the \textsc{Rockstar} halo finder, where particles are tagged based on whether or not they are bound to multiple host haloes based on a boundedness criterion described in \citet{Behroozi2012a}. Following the removal of subhalo particles, we recomputed the mass profile of the diffuse dark matter halo and used those as output when re-training the GBT. The ICs and MAH features remained identical to the original case where all particles are used to compute the mass profiles.

\begin{figure}
    \centering
	\includegraphics[width=\columnwidth]{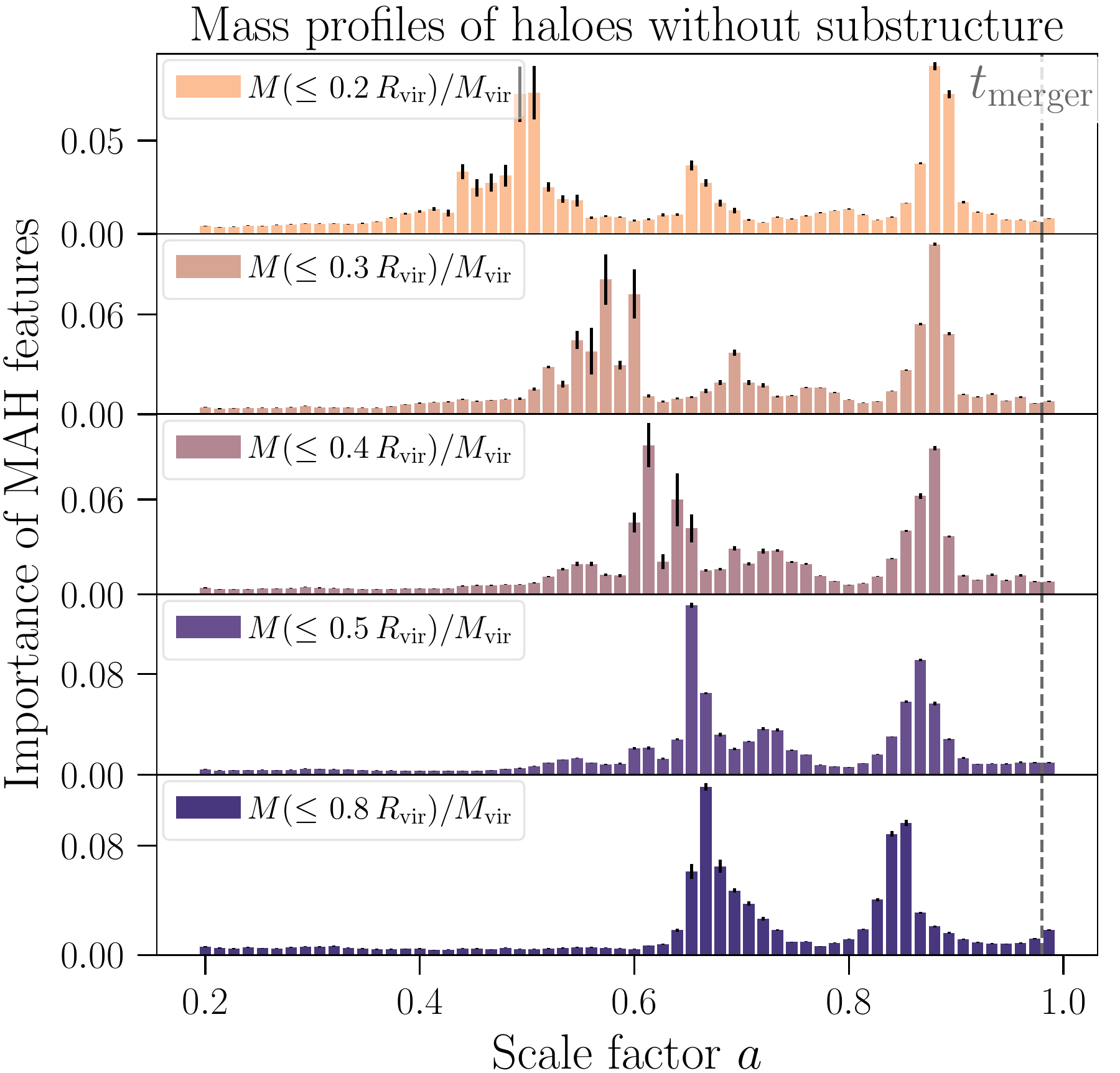}
    \caption{Importances of the MAH features when the GBTs are trained to predict the mass profiles of haloes without any substructures. The third, late-time peak in the MAH importances found in Fig.~\ref{fig:icshistimp} has disappeared. This confirms our interpretation that the third MAH importances peak is related to the impact of recently accreted substructures on the halo mass profile. The first and second peak, related to the virialization time and the dynamical time, remain important for determining the mass profile of the diffuse matter component of the haloes.}
    \label{fig:imp_no_sub}
\end{figure}

Figure \ref{fig:imp_no_sub} shows the importances of the MAH features returned by the GBTs when trained to predict the mass profile of haloes without any substructures. By comparing the importances of the MAH as we train the model on haloes with and without substructure, we can directly test whether any relevant time-scales found in the former case are related to the impact of massive substructures on the final mass profiles. We find that the latest time peak found in Fig.~\ref{fig:icshistimp} has disappeared; this suggests that the MAH of haloes at very late times is relevant to capture the impact of substructures on the mass profile, as already mentioned in Sec.~\ref{sec:peakthree}. This is expected since cluster-mass haloes are the youngest objects in the Universe, containing massive substructures that have only recently been accreted onto the halo and not yet destroyed. Therefore, the GBT correctly identifies the late-time accretion of these massive substructures to be a relevant feature for correctly describing the outer profile of clusters. The fact that the last peak largely disappears when substructure is removed, implies that a lot of the material on first infall is brought in by massive substructure.

As mentioned in Sec.~\ref{sec:impmahs}, the first peak at $a\sim 0.6$ corresponds to the formation time of the halo: the mass distribution of a halo is affected by the time at which it accreted its mass and became virialized. Note that compared to Fig.~\ref{fig:icshistimp}, the first peak in the MAH importances shows a gradual shift to earlier times; this shows that the accretion and disruption of subhaloes at late times affects the estimate of the formation time of the halo at different radial scales.
The second, late-time, peak in the MAH importances in Fig.~\ref{fig:imp_no_sub} is clearly recovered when the GBTs are trained to predict the profile of the diffuse dark matter component. This is particularly true for the inner radial bins. In the innermost radial bin (top panel in Fig. ~\ref{fig:imp_no_sub}), the late time peak shows that the mass within that radius is significantly affected by material accreted $\sim 1$ Gyr back; this is the pericentre crossing time for recently accreted matter. Removing particles that are bound to substructure makes this connection clearer. As particles travel on their orbits past pericentre, towards splashback, they move to larger radial bins making the late time peak shift to earlier scale factors for the outer radial bins. The late-time peak corresponds to material between pericentre and the splashback radius. This material is dynamically unrelaxed and affects the profile on all scales.

In summary, this test shows the connection between the third, most recent peak in the MAH importances (Fig.~\ref{fig:icshistimp}) and the late-time accretion of massive substructures affecting the halo mass profile especially in the outskirts. The first two peaks, related to the virialization time and the dynamical time respectively, remain important for determining the diffuse mass profile of haloes at all radii.

\subsection{MAH importances peaks and particles' accretion times}
\label{sec:accparticles}
\begin{figure}
    \centering
	\includegraphics[width=\columnwidth]{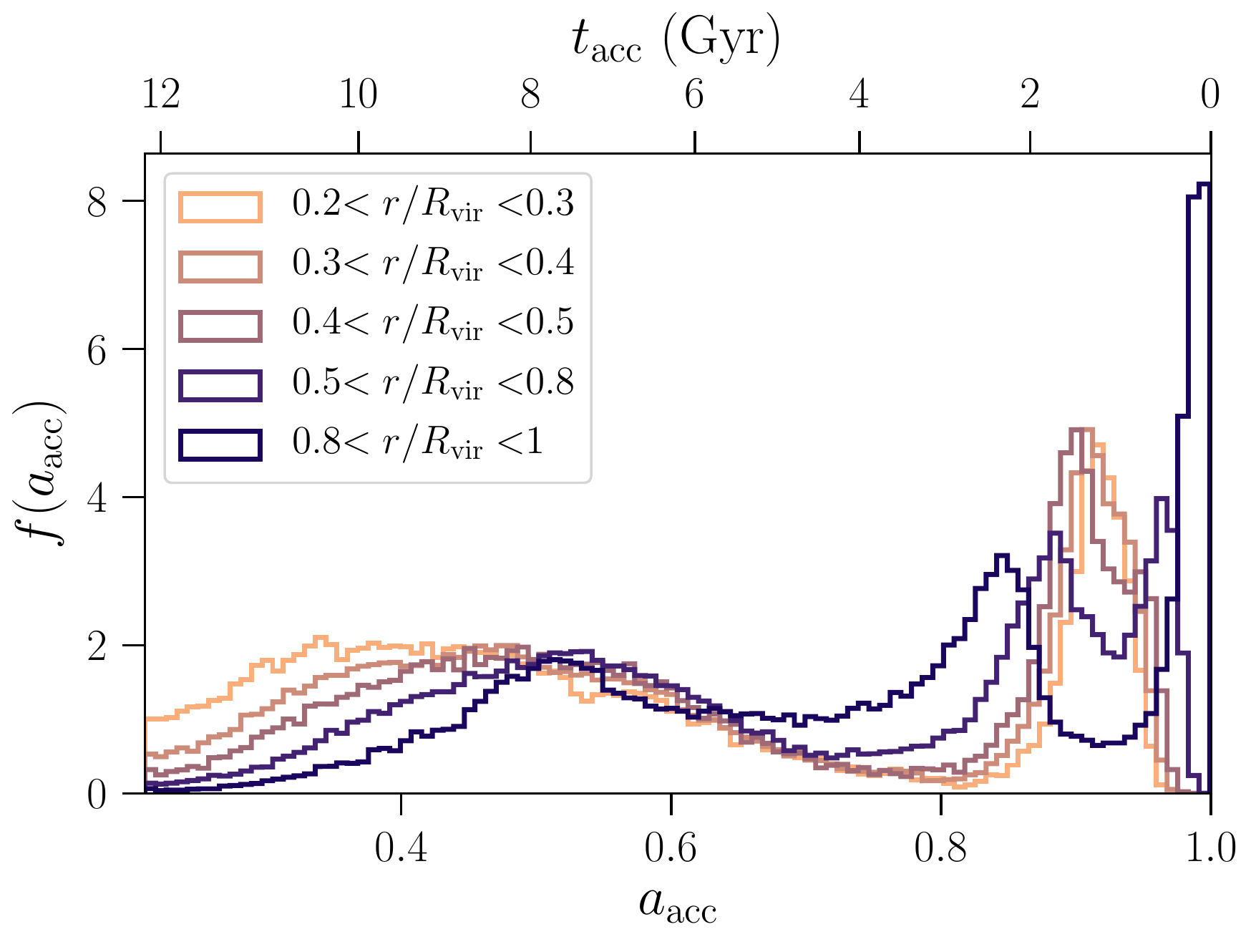}
    \caption{An example of distribution of scale factors, $a_{\rm acc}$, at which particles were accreted within $R_{\rm vir}$ of an individual cluster mass halo. The top axis shows the lookback time at accretion. The different histograms show the accretion times of particles located at different radii within the halo at $z=0$. The peaks in the accretion time distributions reflect the peaks in the MAH importances revealed by the GBT.}
    \label{fig:infalltimes}
\end{figure}

The separation of time-scales that we observe in the importances of the MAH reflects the composition of different parts of the halo. To build a qualitative intuition of this connection, we show in Fig.~\ref{fig:infalltimes} the distribution of accretion times of dark matter particles into an individual cluster host halo. We isolate all the particles within the virial radius of the halo and trace the orbit of each particle back in time to $a=0.2$. We evaluate the time at which the particle crosses into the virial radius of the host for the first time, $t_{\rm acc}$. We plot the distribution of accretion times for particles in different radial shells around the host at $z=0$. We find that the distribution of infall times of particles at different distances show peaks at locations similar to those in the importances of the MAHs. 

We find that at all radii, the dark matter particles have two distinct populations: a set of particles that have been accreted at early times ($a_\mathrm{acc} \lesssim 0.8$) that is most likely phase-mixed and can be thought of as the relaxed material, and a second set of particles that has been more recently accreted ($a_\mathrm{acc} \gtrsim 0.8$). In the outer regions of the cluster (darker histograms in Fig.~\ref{fig:infalltimes}), we in fact see three different populations: the older particles ($a_\mathrm{acc} \lesssim 0.8$), the particles that have completed one pericentre and are approaching that radius for the second time in their orbits ($a_\mathrm{acc} \sim 0.85$) and the particles that are crossing that radius for the first time falling in ($a_\mathrm{acc} \sim 0.98$). The peaks of the accretion times closely resemble the peaks of the MAH importances shown in the $M(\leq 0.8 \, \rvir{})/\Mvir{}$ panel of Fig.~\ref{fig:icshistimp}. The qualitative connection between accretion times and the MAH importances hints at the presence of a dynamical population of particles in the halo that retains a distinct signature on the final mass profile \citep{Adhikari2021, Diemer2022}.

\section{Comparison with linear models}
\label{sec:linear}
To better understand the nature of the impact of the various input features on the mass profile, we compare the results of the GBT with a simple linear regression model. The latter takes the same inputs as the GBT and fits a linear model to minimize the residual sum of squares between the ground truths $M(\leq r)/\Mvir{}$ and the values predicted by the linear approximation. Similar to the GBT, we use a separate linear regression model to infer the values of $M(<r)/\Mvir{}$ for $r \in \{0.2, 0.3, 0.4, 0.5, 0.8, 3.0\} \rvir{}$, respectively.

\begin{figure}
    \centering
	\includegraphics[width=\columnwidth]{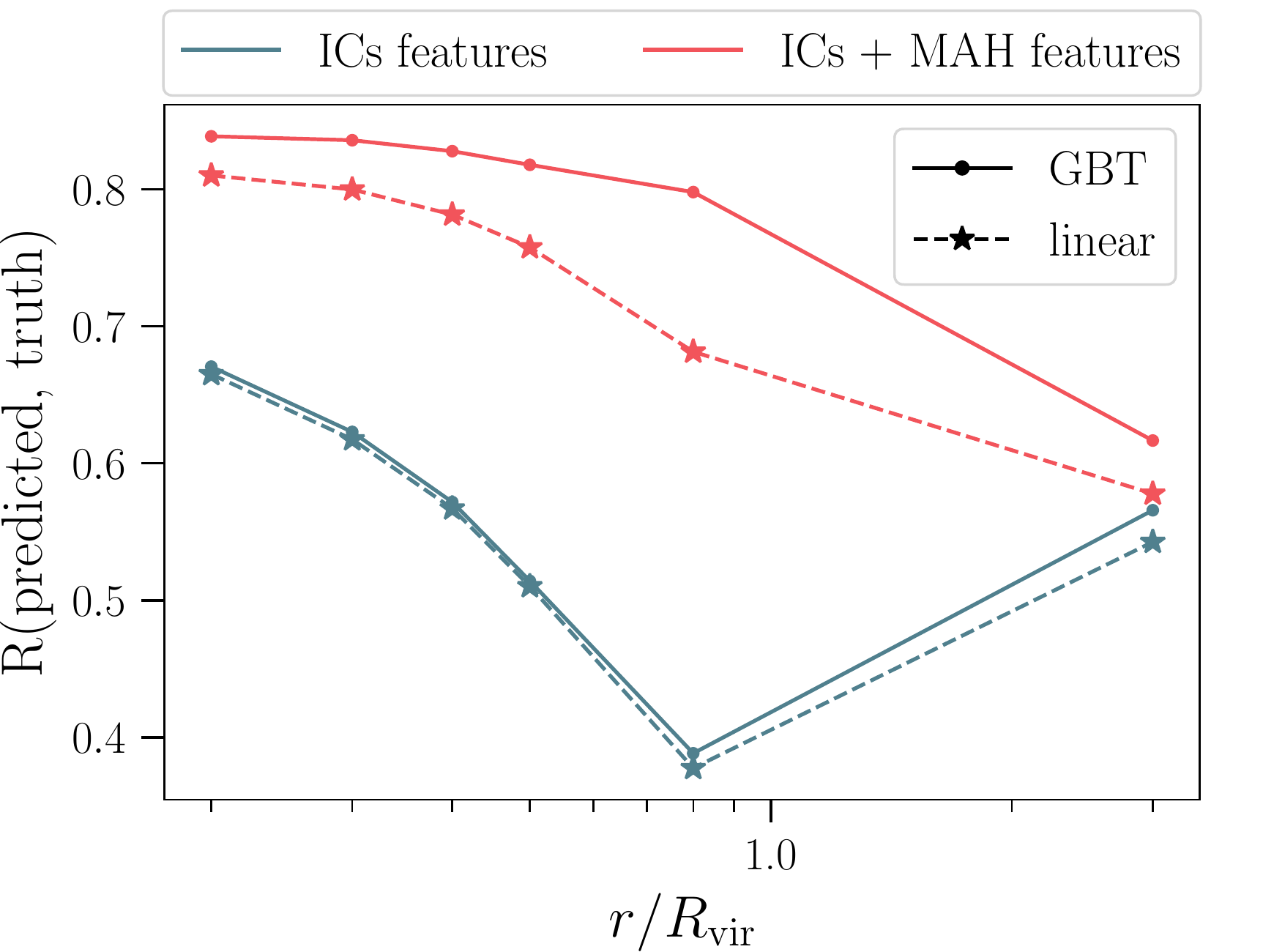}
    \caption{Correlation coefficient $R$ between predicted and ground truth values of $M(\leq r)/\Mvir{}$ as a function of $r/\rvir{}$ for the GBT and a linear regression model. The response of the mass profile to the ICs smoothed overdensities is linear, whereas the profile response to the MAH is characterized by a non-linear relationship.}
    \label{fig:linear_vs_GBT}
\end{figure}

Figure~\ref{fig:linear_vs_GBT} shows the correlation coefficient between predicted and ground truth values of $M(\leq r)/\Mvir{}$ as a function of $r/\rvir{}$ for the GBT and linear models. We find similar predictive accuracy when both the GBT and the linear model are trained based on the ICs features alone. This suggests that the response of the mass profile to the ICs smoothed overdensities is linear. On the other hand, the GBT outperforms the accuracy of the linear models as we add the MAH features to the inputs. This suggests that the response to the MAH is characterized by a non-linear relationship. 

\section{Discussion}
\label{sec:discussion}
In this work, we first trained the GBTs to infer the final mass profile of cluster-mass haloes based on information about the ICs alone. We find that there are two primary scales in the ICs density field that affect the final mass profile (Fig.~\ref{fig:icsimp}). The first is given by the density near the Lagrangian radius of the haloes ($R\sim 5~\mathrm{Mpc}/h \sim 0.7~R_L$) and primarily affects the inner profile; the second is the density in the large-scale environment ($R\sim 12~\mathrm{Mpc}/h \sim 1.7~R_L$) which becomes increasingly important in predicting the final profile towards the outskirts. This result confirms the link between large-scale environment in the ICs and the haloes' outer profile. We test the robustness of our results for a different mass range of cluster-mass haloes, which also revealed that the most relevant information lies in the initial density field smoothed on $R\sim 0.7~R_L$ and $R\sim 1.6~R_L$. 

In terms of predictive accuracy, the model is capable of predicting the mass in the inner profile based on information about the initial density alone to reasonable accuracy (Fig.~\ref{fig:icspredictions}). The correlation coefficient between predicted and true values of $M(<r)$ is $\sim 0.7$ in the inner profile. This suggests that the inner profile retains memory of the ICs. On the other hand, the predictive accuracy of the model decreases as we move towards the virial boundary of the haloes, suggesting that there is additional information required, beyond the ICs spherical overdensities, to correctly predict the enclosed mass in the outskirts.

As we add information about the haloes' MAHs, the final mass profile can be predicted with significantly improved accuracy at all radii (Fig.~\ref{fig:icshistpredictions}). The initial density field smoothed over the haloes' Lagrangian scale and the early-time accretion phase equally contribute to the build up of the haloes' inner profile, while the later-time mass assembly history dominates over information stored in the ICs when determining the outer profile (Fig.~\ref{fig:icshistimp}).

\begin{figure}
    \centering
	\includegraphics[width=\columnwidth]{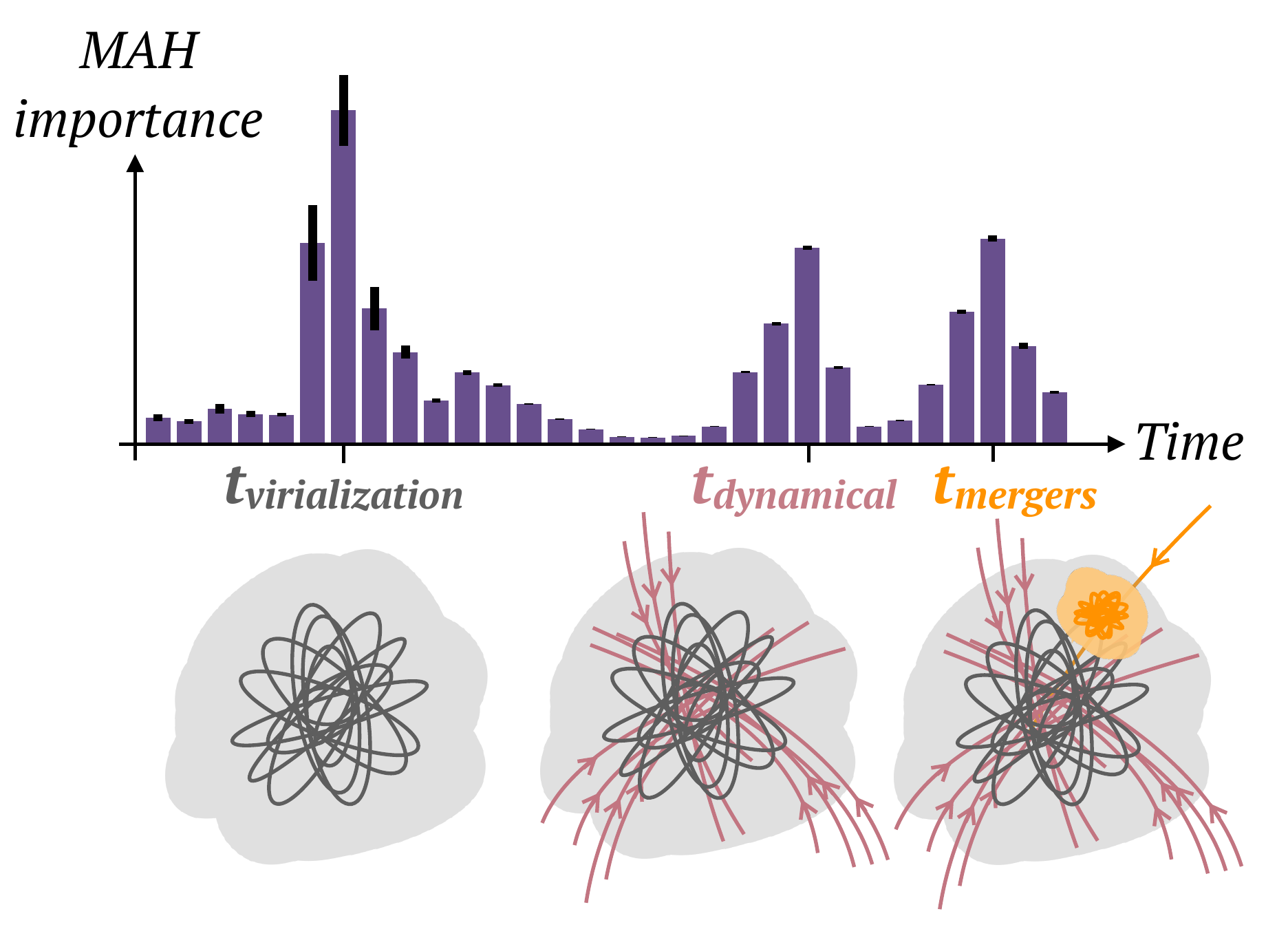}
    \caption{Illustration of the physical interpretation of the most important time-scales in the MAH of haloes discovered by the machine-learning model. The innermost peak corresponds to particles accreted at early times that are in tight orbits (shown in grey), the second peak corresponds to particles on their first orbits, out to splashback, that are dynamically unrelaxed (shown in red) and the third peak corresponds to subhaloes on recent infall (shown in orange).}
    \label{fig:illustration}
\end{figure}

We find that there are three time-scales in the MAHs that carry the most relevant information needed to accurately predict the final mass profiles (Fig.~\ref{fig:icshistimp}). The first is related to the formation time of the haloes ($a\sim0.6$): the time at which mass is first assembled into a virialized halo impacts the final mass distribution at all radii. The second relevant time-scale is likely related to the dynamical time ($a\sim0.85$): this time-scale dictates how the final mass distribution is affected by the dynamically unrelaxed, infalling material inside the halo. The third time-scale appears at very recent times in the haloes' assembly history ($a\sim0.98$): this captures the impact of massive substructures that have recently merged with the haloes on the outer profile. Fig.~\ref{fig:illustration} illustrates the physical interpretation of the three most important time-scales in the MAH identified by the GBTs. 

Our result suggests that the virialized and dynamically unrelaxed component of haloes need to be modelled separately, particularly for cluster mass objects. This is similar to the results of \citet{Diemer2022}, where the virialized and first infall (upto pericentre) components are separated (see also \citealt{Garcia_inprep}). Our work suggests that dark matter out to its first orbit apocentre, i.e. nearly to the epoch when it reaches splashback radius, remains unmixed in phase and is separate from the virialized component, forming what can be thought of as the halo's quiet interior \citep{More:2015ufa}. These time-scales are also consistent with those observed in \cite{Valluri:2006vi}. Our results also agree with those in \citet{Ludlow2012}: they show that dynamically unrelaxed haloes with recently accreted material in the outskirts have high concentrations, contrary to the typical expectation that high-concentrated haloes form at earlier times. Their results support our finding that the dynamically unrelaxed component of haloes plays a major role in determining the final halo profile, and that it must be modelled separately from the virialized component.

\section{Conclusions}
\label{sec:conclusions}

We have presented a machine-learning framework to provide physical insights into the origin of the final mass profiles of cluster-sized haloes. We used gradient-boosted-trees to predict the final mass profile of a halo given two sets of inputs: (1) the smoothed density field in the ICs around the centre-of-mass of the halo's Lagrangian patch and (2) the halo's MAH. The strength of our approach lies in its ability to establish a physical interpretation of the machine-learning results. We quantified the feature importances of the different inputs during the training process of the algorithm, thus revealing the profile's sensitivity to the ICs and the haloes' mass assembly histories.

We find that the final mass profiles retain memory of the ICs, especially in the haloes' inner region. The shape of the profile is sensitive to the ICs spherical overdensity near the Lagrangian patch size, and in the large-scale environment. This result confirms the presence of a link between the environment in the ICs and final profile. 

The halo's MAH contributes to the mass profile at all radii, in particular erasing memory of the ICs spherical overdensities in the halo outskirts. We show that the halo mass profile consists of different components that were accreted at three separate time-scales in the MAH. We find that the profile at all radii is sensitive to matter that was accreted around the formation time of the halo; this forms the virialized or relaxed component of the halo. Additionally, the mass profile is affected by (i) matter on its first orbit accreted over the last dynamical time, and (ii) recently accreted material (before pericentre crossing) residing primarily within subhaloes. This dynamical component of the halo contributes to the mass profile at all radii and its contribution is comparable to that of the virialized component of the halo, particularly in the outskirts. Our results confirm the presence of a virialized and a dynamically unrelaxed component of the haloes, and reveal the components' connection to specific time-scales in the haloes' MAH.

Our work demonstrates the utility of interpretable machine-learning techniques to gain physical understanding of large-scale structure formation. Our approach relies on the ability to establish a physical interpretation of the outputs of the machine-learning model. In future work, we plan to extend our framework to investigate the origin of assembly bias and the mass profile of lower mass haloes, and to further adopt deep learning techniques to extract unknown features of the dataset that are relevant to cosmological structure formation.

\section*{Author Contributions}
\textbf{L.L.-S.}: project conceptualization; formal analysis; methodology; software; investigation, validation \& interpretation; writing - original draft, editing, final; visualisation; funding acquisition.
\textbf{S.A.}: project conceptualization; formal analysis; methodology; software; investigation, validation \& interpretation; writing - editing; funding acquisition.
\textbf{R.H.W.}: project conceptualization; methodology; investigation, validation \& interpretation; writing - editing; funding acquisition.

\section*{Acknowledgements}
We thank Arka Banerjee, Peter Behroozi, Corentin Cadiou, Neal Dalal, Sten Delos, Benedikt Diemer, Yao-Yuan Mao, Marcello Musso, Ethan Nadler, Hiranya Peiris and Andrew Pontzen for useful discussions. We also thank Andrew Pontzen for providing feedback on the paper draft. This work was supported by collaborative visits funded by the Cosmology and Astroparticle Student and Postdoc Exchange Network (CASPEN), and received additional support from the Kavli Institute for Particle Astrophysics and Cosmology (KIPAC) at Stanford and SLAC National Accelerator Laboratory and from the U.S. Department of Energy under contract number DE-AC02-76SF00515 to SLAC National Accelerator Laboratory.
LLS acknowledges the hospitality of KIPAC, where part of this work was completed. This project has received funding from the European Union's Horizon 2020 research and innovation programme under grant agreement No. 818085 GMGalaxies. SA was partially supported by the U.S. Department of Energy (DOE) Office of Science Distinguished Scientist Fellow Program. This work used computing facilities provided by the UCL Cosmoparticle Initiative, SLAC National Accelerator Laboratory, a U.S. Department of Energy Office, and the Sherlock cluster at the Stanford Research Computing Center (SRCC); we are grateful to the computational support teams at each of these centres. 

\section*{Data Availability}
The data underlying this article will be shared upon reasonable request to the corresponding author.


\bibliographystyle{mnras}
\bibliography{main}



\appendix
\section{Splitting haloes by their concentration}
\label{sec:conc}

We investigate how the importances vary if we split the haloes based on their final concentration. We split the halo population across 10 different halo mass bins, and for each bin we split the haloes into two populations with concentration above and below the median concentration. We then trained (and tested) the two populations separately, and compared the importances of the features for the two cases.

\begin{figure}
    \centering
	\includegraphics[width=\columnwidth]{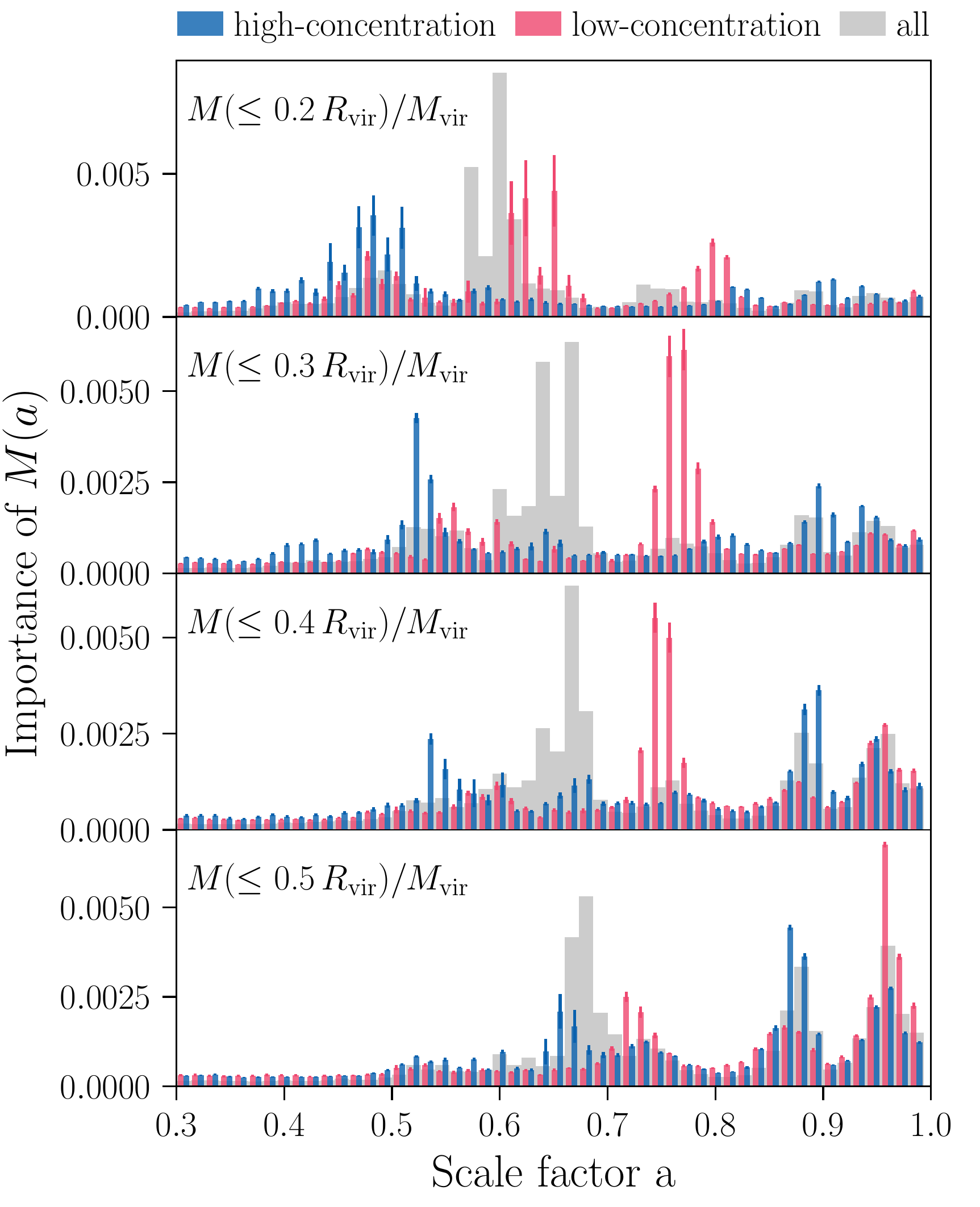}
    \caption{Importances of the MAH features when the GBTs are trained separately on haloes with higher (blue) or lower (pink) concentration than the median concentration. The grey histogram shows the MAH feature importances when the GBTs are trained on all haloes, for reference. The GBTs recover the expected trend: more concentrated haloes tend to form earlier compared to less concentrated haloes which form later.}
    \label{fig:splitc_hist_imp}
\end{figure}

Figure \ref{fig:splitc_hist_imp} shows the importances of the MAH features when the GBT models are trained (and tested) on haloes with high concentration (blue) and low concentration (pink), separately. The different panels show the importances of the features in determining the profile at $r=0.2\, \rvir{}$ up to $r=0.5\, \rvir{}$. At larger radii, we find no differences between the importances of the two populations, hence we do not show those in Fig.~\ref{fig:splitc_hist_imp}. Comparing the two histograms shows how the relevance of MAH changes when determining the mass profile of high concentration vs low concentration haloes. The two histograms for high-c and low-c haloes follow the expected trends: more concentrated haloes tend to form earlier compared to less concentrated haloes which form later. Therefore, the MAH importances for more concentrated haloes peak at a smaller scale factor (earlier times) than for less concentrated haloes. On the other hand, the feature importances for the case of the entire halo population peak at a scale factor that is in between the peaks of the high-concentration and low-concentration samples. The two late-time peaks in the importances of whole population of haloes are found also for the two populations of high-concentrated and low-concentrated haloes. This suggests that late-time accretion affects the profile of the two halo populations in similar ways.

\bsp	
\label{lastpage}
\end{document}